\documentclass[seceq]{ptptex}

\usepackage{graphicx}
\def\tr{\mathop{\rm tr}\nolimits}

\def\Re{\mathop{\rm Re}\nolimits}
\def\Im{\mathop{\rm Im}\nolimits}
\def\Log{\mathop{\rm Log}\nolimits}

\newcommand{\ol}{\overline}

\title{%        %You can use \\ for explicit line-break
Exactly Marginal Deformations of
Quiver Gauge Theories as Seen from Brane Tilings%
}

\author{%       %Use \scshape  for the family name
Yosuke \textsc{Imamura},\thanks{E-mail: imamura@hep-th.phys.s.u-tokyo.ac.jp}\quad
Hiroshi \textsc{Isono},\thanks{E-mail: isono@hep-th.phys.s.u-tokyo.ac.jp}\quad
Keisuke \textsc{Kimura},\thanks{E-mail: kimura@hep-th.phys.s.u-tokyo.ac.jp}\\
and
Masahito \textsc{Yamazaki}\thanks{E-mail: yamazaki@hep-th.phys.s.u-tokyo.ac.jp}%
}

\inst{%         %Affiliation, neglected when [addenda] or [errata]
Department of Physics, The University of Tokyo,
Tokyo 113-0033, Japan}

\recdate{February 16, 2007}%            %Editorial Office will fill in this.

\abst{%         %this abstract is neglected when [addenda] or [errata]
We study the relation
between
exactly marginal deformations in a large class of
${\cal N}=1$ superconformal quiver gauge theories
described by brane tilings
and  the degrees of freedom in the corresponding
$5$-brane systems.
We show, with the help of NSVZ exact $\beta$ functions,
that there are generically $d-1$ complex exactly marginal deformations
of a gauge theory, where $d$ is the perimeter
of the corresponding toric diagram.
We identify $d-3$ complex marginal deformations
as deformations of the brane system
and the Wilson lines on it,
and the other two,
the diagonal gauge coupling and a $\beta$-like deformation,
as background supergravity fields.
}

\begin{document}

\maketitle

%%%%%%%%%%%%%%%%%%%%%%%%%%%%%%%%%%%%%%%%%%%%%%%%%%%%%%%%%%%%
\section{Introduction}
Brane tilings\cite{Hanany:2005ve,Franco:2005rj,Franco:2005sm,Hanany:2005ss,Feng:2005gw,Franco:2006gc,Hanany:2006nm,Garcia-Etxebarria:2006aq,Butti:2006hc,Oota:2006eg,Lee:2006hw,Butti:2005ps,Butti:2005vn,Butti:2006nk,Butti:2005sw}
are useful tools to describe a large class of
${\cal N}=1$ superconformal quiver gauge theories. 
They are
simply bipartite graphs (i.e., graphs consisting of two kinds of vertices, usually colored black and white, and links connecting vertices of different colors) on tori. Bipartite graphs are dual
to quiver diagrams on tori (known as periodic quiver diagrams).
Specifically, the faces (resp., edges) in a brane tiling
correspond to the $SU(N)$ gauge groups (resp., bi-fundamental chiral multiplets) of the corresponding quiver gauge theory.

Brane tilings have several advantages over conventional quiver diagrams.
For example, they clarify the relation between quiver gauge theories and
the geometric structure of
the three-dimensional toric Calabi-Yau cones,
as there is a simple method to obtain toric diagrams, which encode
all the geometric information of the cones,
from bipartite graphs.\cite{Hanany:2005ve,Franco:2005rj} \ 
Conversely, given a toric diagram, we can construct the brane
tiling\cite{Hanany:2005ss}, which in turn gives us the quiver and
the superpotential (up to some ambiguities which are
conjectured\cite{Beasley:2001zp,Hanany:2005ss} to be nothing but the Seiberg duality\cite{Seiberg:1994pq}). The validity of these methods has been confirmed in numerous examples, and some parts of them have been rigorously proved valid.\cite{Franco:2006gc,UY}

For the reasons given above, we can regard brane tilings as bridges between
geometries and gauge theories, and they are useful
for studying Calabi-Yau geometry itself.\cite{Feng:2005gw} \ For example,
in Refs.~\citen{UY,UY2,UY3} it was shown that brane tilings are useful in proving a
version of mirror symmetry known as the homological mirror symmetry.

Another virtue of brane tilings is that they actually represent {\it physical} brane systems. Although originally introduced as technical tools in combinatorics, their physical meaning was subsequently clarified in Ref.~\citen{Feng:2005gw}. Explicitly, we can regard brane tilings as physical D5/NS5-brane systems on which the quiver gauge theories are defined. 
The faces are then interpreted as stacks of D5-branes, and the links are interpreted as the intersections of the D5-branes with NS5-branes.

Some works have been motivated by this understanding. For example,
in Ref.~\citen{Imamura:2006ub} it is shown that
the cancellation conditions of gauge and $U(1)_B$ anomalies are
reproduced as boundary conditions imposed on fields on branes
at the intersections of
the D5-branes and NS5-branes.
We can also reproduce 't Hooft anomalies as
anomaly inflows in D5-NS5 systems.\cite{Imamura:2006ie} \
For these analyses, only the topological structure of the brane systems
is relevant, and we do not have to consider the actual shapes of
branes which satisfy equations of motion.

It is natural to ask to what extent we can derive the
properties of gauge theories from brane tilings treated as representing real brane systems, rather than just as a combinatorial and topological tool.

If we take account of the equations of motion, or the equivalent
BPS conditions, we can no longer use bipartite graphs constructed
on flat tori as the real shapes of brane configurations.
The D5-brane worldvolume on which a diagram is constructed is bent by the tension
of the NS5-branes, and its shape is determined by the equations of motion. 
Furthermore, the bipartite graphs do not even give the topological structure of
brane systems.
In bipartite graphs,
all faces are regarded as stacks of D5-branes.
In brane systems, by contrast,
in addition to pure D5-branes and NS5-branes,
bound states of D5-branes and NS5-branes appear,
and so we have a brane system consisting of more than two types of branes.

The problems which we address in this paper are
to determine the number of parameters possessed by the brane configurations
and the correspondence between these parameters and those in gauge theories.

We first investigate exactly marginal deformations of quiver gauge theories
in \S\ref{gauge.sec}. This is done by analyzing $\beta$-functions for gauge couplings and superpotential couplings.
The analysis of brane systems is carried out in \S\ref{brane.sec}.
For the purpose of constructing gauge theories, we need to take the
decoupling limit, with $g_{\rm str}\rightarrow0$, to decouple
unwanted modes.
In this limit, the NS5-brane worldvolume becomes a holomorphic
curve in $({\mathbb C}^*)^2$ and is described by a Newton polynomial.
The D5-brane worldvolumes are disks whose boundaries are on the holomorphic curve.
We show that the BPS conditions for D5-branes impose
some conditions on the coefficients in the Newton polynomial.
We determine the dimensions of parameter spaces for brane systems.
In simple examples, we can solve the BPS conditions and
explicitly derive the constraints imposed on the coefficients.
It is, however, difficult to solve the BPS conditions in general cases.

We also analyze the brane system in the opposite limit, $g_{\rm str}\rightarrow\infty$.
Even though the relation to the gauge theories in this limit is
not clear, the shape of the brane configuration becomes quite simple,
and we can solve the BPS conditions
for general tilings.

The results of the gauge theory analysis and brane analysis are
compared in \S\ref{comparison.sec},
and we propose a correspondence between the sets of
deformation parameters in the two cases.

In the rest of this section,
we briefly explain the method for constructing the brane configurations.
The relation between gauge theories and the
topological structures of brane configurations
is quite useful, even for analysis carried out in gauge theories.

Let us consider ten-dimensional
spacetime with two directions, $x^5$ and $x^7$,
compactified on ${\mathbb T}^2$.
We start the construction with a stack of $N$ D5-branes wrapped on
this ${\mathbb T}^2$
(see Table \ref{config.tbl}).
\begin{table}[t]
\caption{The brane configuration corresponding to brane tilings. The directions 5 and 7 are compactified, and $\Sigma$ is a two-dimensional surface in 4567 space.}
\label{config.tbl}
\begin{center}
\begin{tabular}{ccccc|cccc|cc}
\hline\hline
& 0 & 1 & 2 & 3 & 4 & 5 & 6 & 7 & 8 & 9 \\
\hline
NS5 & $\circ$ & $\circ$ & $\circ$ & $\circ$ &\multicolumn{4}{c|}{$\Sigma$} & & \\
D5  & $\circ$ & $\circ$ & $\circ$ & $\circ$ && $\circ$ && $\circ$ & &\\
\hline
\end{tabular}
\end{center}
\end{table}
If ${\mathbb T}^2$ is sufficiently small and
the Kaluza-Klein modes decouple,
we have four-dimensional ${\cal N}=4$ $SU(N)$ super Yang-Mills theory
on the D5-branes.
(We ignore the diagonal $U(1)$ here.
However, we note that it plays an important role when we consider baryonic $U(1)$
global symmetries.\cite{Imamura:2006ub,Imamura:2006ie})
In terms of ${\cal N}=1$ multiplets, this theory consists of
an $SU(N)$ vector multiplet and three adjoint chiral multiplets.

For the purpose of obtaining multiple $SU(N)$,
we divide this worldvolume into
faces by attaching NS5-branes wrapped on $1$-cycles in ${\mathbb T}^2$.
Let $d$ be the number of NS5-branes.
We label them as $\mu=1,\ldots,d$
and denote by $\alpha_\mu$ the cycle on which the $\mu$-th NS5-brane is wrapped.
In $4567$ space, their worldvolumes are semi-infinite cylinders
with boundaries $\alpha_\mu$ on the D5-brane worldvolume.
At the boundaries, the NS5 charge flows into or out of ${\mathbb T}^2$,
and the NS5 charge on ${\mathbb T}^2$ jumps by $\pm1$ on the cycles.
As a result, ${\mathbb T}^2$ is no longer contains only D5-branes but now also
contains $(N,k)$-branes, the bound states of $N$ D5-branes and $k$ NS5-branes.

There may in general exist faces with $|k|\geq2$,
and in \S\ref{strong.sec}
we actually find that in strong coupling limit,
$g_{\rm str}\rightarrow\infty$, there is an example with such faces.
In the weak coupling limit, however, we never have such faces,
because the NS5-brane is wrapped on a holomorphic cycle,
as discussed below.
Thus, in this case, we have $(N,0)$ and $(N,\pm1)$-branes.

The boundaries of NS5-branes divide ${\mathbb T}^2$ into many faces.
This partition changes the gauge group and the matter content.
Because the $SU(N)$ gauge fields can exist only on $(N,0)$-branes,
we have one $SU(N)$ gauge group for each $(N,0)$ face.
The three adjoint chiral multiplets we had at the beginning
disappear, because the attached NS5-branes fix the position of the
D5-branes.
Instead, we have new chiral multiplets at the intersections of cycles
at which two $(N,0)$ faces contact each other.
A bi-fundamental chiral multiplet arises at each intersection
as massless modes of
open strings ending on these two faces.
As a result, we obtain ${\cal N}=1$ quiver gauge theories.

When the NS5 charges of all faces are $0$ or $\pm1$,
we obtain a bipartite graph from the partition
by replacing the $(N,+1)$ and $(N,-1)$ faces
by black and white vertices, respectively,
and connecting them by links according to adjacency of the faces.
Conversely, we can also easily obtain the partition of a torus from a bipartite graph.\cite{Hanany:2005ss} \
In the literature, brane tilings are usually described by bipartite graphs
instead of the partitions mentioned above.
In this paper, however, we use the term ``brane tilings'' to refer to the partitions of tori into
$(N,0)$ and $(N,\pm1)$ faces.

\begin{figure}[t]
\centerline{\includegraphics[scale=0.5]{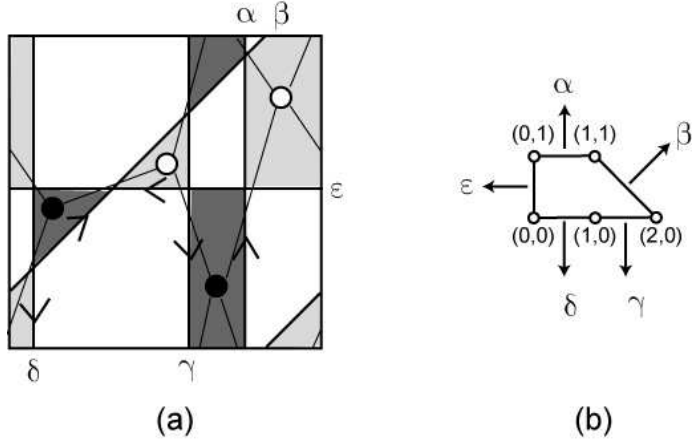}}
\caption{An example of a partition of the D5 worldvolume (a)
and
the corresponding toric diagram (b).}
\label{sppima.eps}
\end{figure}

Figure \ref{sppima.eps} (a) depicts an example of the partition
with the corresponding bipartite graph superposed.
The five NS5 boundaries, $\alpha$, $\beta$, $\gamma$, $\delta$ and $\epsilon$,
divide the torus into seven faces.
The three white faces are $(N,0)$-branes
and give the gauge group $SU(N)^3$.
The two light and two dark shaded faces are $(N,-1)$-branes and
$(N,1)$-branes, respectively.
We have seven chiral multiplets, corresponding to seven intersections of
cycles.

The cycles in the tiling can be identified with the
normal vectors of edges in the toric diagram
of the dual Calabi-Yau geometry [(b) in Fig.~\ref{sppima.eps}].
In this example, the geometry described by the toric diagram (b) in
Fig.~\ref{sppima.eps} is the suspended pinch point (SPP).

%%%%%%%%%%%%%%%%%%%%%%%%%%%%%%%%%%%%%%%%%%%%%%%%%%%%%%%
\section{Exactly marginal deformations}\label{gauge.sec}
In this section, we study exactly marginal deformations
of gauge theories described by brane tilings, along the lines of
Ref.~\citen{Leigh:1995ep}.
Even though the analysis in this section is field theoretical,
and we do not use information from string theory,
brane tilings are still useful as diagrams representing
the structure of gauge theories.

In this paper,
we only consider a superpotential of the form
\begin{equation}
W=\sum_k\pm h_k\tr\prod_{I\in k}\Phi_I,
\label{hkterm}
\end{equation}
where $k$ and $I$ label $(N,\pm1)$ faces in the brane tiling
and intersections of cycles, respectively.
The signature is chosen according to the NS5 charge of the
corresponding face.
The coefficients $h_k$ are coupling constants, and
$\Phi_I$ is the bi-fundamental field corresponding to the
intersection $I$.
The expression $I \in k$ means that $I$ is one of the corners of
the face $k$.
In the following, we use $k$ to label $(N,\pm1)$ faces, while $a$ is used for
$(N,0)$ faces.
When we use the indices $A$, $B$, $\ldots$, they run over all the faces.
In general, additional ``accidentally marginal''
operators may appear in the superpotential, and we may have
more degrees of freedom of an exactly marginal deformation.
In this paper, however, we consider only operators appearing in the superpotential
(\ref{hkterm}), which exist generically.

For example, we obtain three complex parameters in the analysis below
for the conifold theory.
It is known, however, that there are five complex parameters
in the theory.\cite{Benvenuti:2005wi} \ 
We do not discuss this point.

The parameters in a gauge theory include
gauge coupling constants for $SU(N)$ gauge groups
and the coefficients in the superpotential.
Conformal fixed points are
defined by the
vanishing of the
beta functions for these parameters.
Let $g_a$ be the gauge coupling of the gauge group $SU(N)_a$ associated
with the $(N,0)$ face $a$.
The beta function for $g_a$ is related to the
anomalous dimensions $\gamma_I$ of the
bi-fundamental fields $\Phi_I$
by the NSVZ exact beta-function formula\cite{Novikov:1983uc}
\begin{equation}
\beta_a\equiv\mu\frac{d}{d\mu}\frac{1}{g_a^2}
=\frac{N}{1-g_a^2 N/8\pi^2}
\left[3-\frac{1}{2}\sum_{I\in a}(1-\gamma_I)\right],
\label{betagauge}
\end{equation}
where the summation is taken over the fields
coupled to the gauge group $SU(N)_a$.
Because the conformal dimension of the field $\Phi_I$
is $1+(1/2)\gamma_I$,
the beta-function of the coefficient $h_k$
in (\ref{hkterm}) is given by
\begin{equation}
\beta_k\equiv\mu\frac{d}{d\mu}h_k=-h_k\left[3-\sum_{I\in k}\left(1+\frac{1}{2}\gamma_I\right)\right].
\label{betapot}
\end{equation}
In order for the theory to be scale independent,
the above beta functions (\ref{betagauge}) and (\ref{betapot})
must vanish:
\begin{equation}
\beta_a=\beta_k=0.
\label{babkz}
\end{equation}
This equation defines the parameter space of the conformal field theory.

Because the number of couplings $g_a$ and $h_k$ is the same as the number of
conditions (\ref{babkz}), in generic situations without supersymmetry,
we have isolated IR fixed points.
In the supersymmetric case, however, both the gauge and
superpotential couplings
(\ref{betagauge}) and
(\ref{betapot}) are given in terms of the anomalous dimensions $\gamma_I$,
and there may be identically vanishing linear combinations of these
$\beta$-functions of the form
\begin{eqnarray}
\beta[S_A]
&\equiv&
\sum_a S_a \frac{1}{N}\beta'_a-\sum_k S_k \frac{\beta_k}{h_k}
\nonumber\\
&=&\sum_a S_a\left[3-\frac{1}{2}\sum_{I\in a}(1-\gamma_I)\right]
+\sum_k S_k\left[3-\sum_{I\in k}\left(1+\frac{1}{2}\gamma_I\right)\right],
\label{sumsbeta}
\end{eqnarray}
with some numerical coefficients $S_a$ and $S_k$.
Instead of $\beta_a$, here we have used
\begin{equation}
\beta'_a=\left(1-\frac{g_a^2N}{8\pi^2}\right)\beta_a
=\mu\frac{d}{d\mu}\frac{1}{g_a^{'2}},
\end{equation}
which is the $\beta$-function for the coupling $1/g^{'2}_a$ defined by
\begin{equation}
d\left(\frac{1}{g^{'2}_a}\right)
=\left(1-\frac{g_a^2N}{8\pi^2}\right)d\left(\frac{1}{g_a^2}\right).
\label{gprime}
\end{equation}

These linear combinations of $\beta$-functions
correspond to RG invariant couplings parameterizing
the orbits of RG flow in the coupling space.
We can assume, at least in the vicinity of the IR fixed manifold,
a one-to-one correspondence between these orbits and points
in the fixed manifold.
Thus, in order to determine what marginal deformations exist
in gauge theories,
we should look for vanishing linear combinations of the form (\ref{sumsbeta}).
This analysis was carried out in Ref.~\citen{Uranga:1998vf} for orbifolded conifolds,
and in Ref.~\citen{Benvenuti:2005wi} for $Y^{p,q}$, including the conifold.
In this work, we generalize these analyses to general quiver gauge theories described by
brane tilings (cf. Ref.~\citen{Wijnholt:2005mp} for analysis from the point of view of exceptional collections). See also Refs.~\citen{Erlich:1999rb} and \citen{Lunin:2005jy}.

We now search for coefficients $S_A$ such that
the linear combination of $\beta$ in (\ref{sumsbeta})
vanishes for any $\gamma_I$.
For the cancellation of the coefficients of $\gamma_I$,
the condition
\begin{equation}
\sum_{a\in I}S_a=\sum_{k\in I}S_k,
\label{sumsasumsk}
\end{equation}
must hold, where the summation on the left-hand side is taken over
gauge groups coupling to the chiral field $I$,
and that on the right-hand side is taken over
terms in the superpotential including the fields $I$.
On the brane tiling, chiral fields correspond to the
intersections of cycles,
and $S_a$ and $S_k$ are numbers assigned to
the $(N,0)$ and $(N,\pm1)$ faces, respectively.
For each intersection, we have two pairs of
faces contacting each other at the intersection,
and the relation (\ref{sumsasumsk}) implies that
the sum of $S_A$ for the two faces in one pair
is the same as that for the two faces in the other pair.
This condition is equivalent to
the existence of numbers $b_\mu$ assigned to cycles
which satisfy the relation
\begin{equation}
S_A-S_B=b_\mu
\label{ssb}
\end{equation}
for two faces $A$ and $B$ adjacent to each other on a cycle $\alpha_\mu$.
In (\ref{ssb}), we assume that when the cycle $\alpha_\mu$ is up-going,
the faces $A$ and $B$ are on the left and right sides of the cycle, respectively.
We call sets of numbers $\{S_A,b_\mu\}$ satisfying the relation
(\ref{ssb}) ``baryonic number assignments'' or ``baryonic charges''.
We can easily show that
the quantities $b_\mu$ must satisfy the relation
\begin{equation}
\sum_{\mu=1}^d b_\mu\alpha_\mu=0
\label{balpha}
\end{equation}
in order for there to exist $S_A$ satisfying the relation (\ref{ssb}).
Because of this relation, the number of independent degrees of freedom
of $b_\mu$ is $d-2$.
If we fix all the $b_\mu$ and one $S_A$, the relation (\ref{ssb}) determines
all the other $S_A$.
Therefore, we have $d-1$ linearly independent baryonic number assignments.

The baryonic number assignments defined above often appear in
the analysis of brane tilings.
For example, they can be used to give anomaly-free rank distributions
of gauge groups.\cite{Benvenuti:2004wx,Butti:2006hc} \
In this paper, we assume that all $SU(N)$ factors in the gauge group
have the same rank, and in this case the gauge anomalies trivially cancel.
If we change the ranks of the gauge groups, however, in general
gauge anomalies arise.
It is known that such gauge anomalies
cancel if we use the
baryonic charge $S_a$ assigned to a face $a$
as the size of the gauge group $N_a$ for the corresponding $SU$
factor.
(Of course, $S_a$ must be a positive integer in this case.)
Another use of baryonic number assignments is
to give anomaly-free baryonic $U(1)$ charges of chiral multiplets.\cite{Benvenuti:2004wx,Butti:2006hc} \
We can identify $S_a$ assigned to $(N,0)$ faces as
baryonic $U(1)$ charges of open string endpoints on the faces\cite{Imamura:2006ie},
and the baryonic $U(1)$ charge of a chiral multiplet $I$
is given by the difference $S_a-S_b$, where $a$ and $b$ are
two $(N,0)$ faces contacting each other at the intersection $I$.
This is why we call sets of numbers $\{S_A,b_\mu\}$
baryonic number assignments.

Now we obtain the condition (\ref{ssb})
for the cancellation of the $\gamma_I$ terms in
(\ref{sumsbeta}),
and if this condition is satisfied, we are left with
\begin{equation}
\beta[S_A]
=3\sum_{\rm faces} S_A -3\sum_{\textrm{intersections}}\bar{S}_I,
\label{sum_beta_general2}
\end{equation}
where we have replaced
the sum over both indices $a$ and $k$
by a sum over all faces in the brane tilings,
and the matter contribution is given by the sum over intersections.
We have also introduced $\bar{S}_I$, defined as half of (\ref{sumsasumsk}):
\begin{equation}
\bar{S}_I=\frac{1}{2}\sum_{a\in I}S_a=\frac{1}{2}\sum_{k\in I}S_k.
\end{equation}

\begin{figure}[t]
\centering
{\small
\begin{tabular}{cc}
\includegraphics[width=60mm]{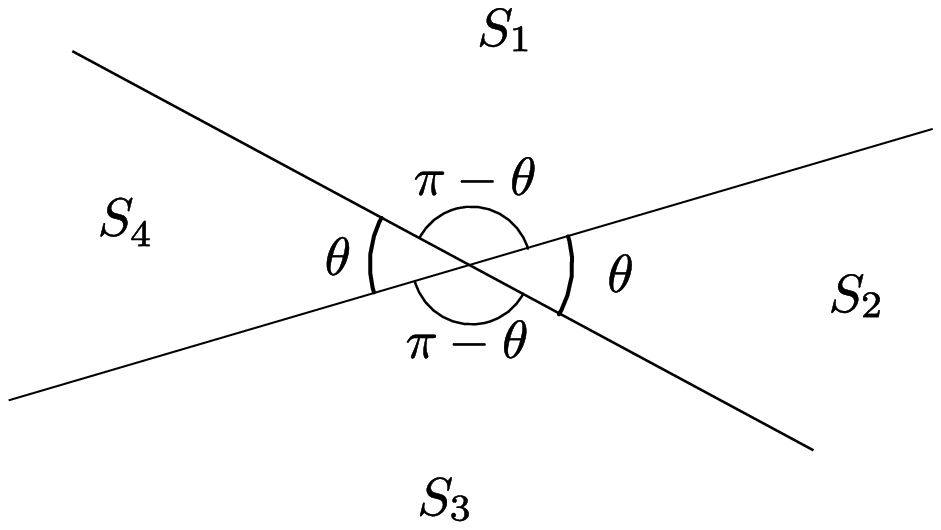}&
\includegraphics[width=60mm]{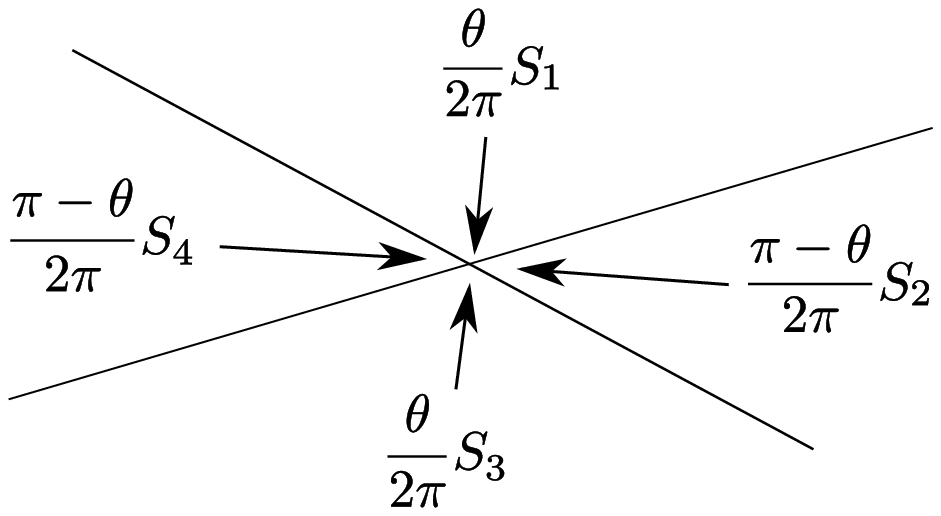}\\
(a) Baryonic charges and angles&
\multicolumn{1}{p{60mm}}{(b) The charges of faces moved to the intersection}
\end{tabular}
}
\caption{Changing the charge distribution of faces to cancel the charges of the intersection.}
\label{s_mean}
\end{figure}

Interestingly, we can show that
if the coefficients $S_A$ satisfy the condition (\ref{ssb}),
the right-hand side of (\ref{sum_beta_general2}) automatically vanishes,
as well as the $\gamma_I$-dependent terms.
To prove this, 
we move a fraction of the total baryonic charge on a given face
equal to the external angle of that face divided by $2\pi$
to the corners of that face.
For example, of
the charge $S_2$ in Fig.~\ref{s_mean},
the amount $[(\pi-\theta)/2\pi]S_2$
is moved to the corner $\theta$.
The sum of the baryonic charges moved to the
corners of one face is
the original baryonic charge of the face,
because the sum of the external angles of a polygon is identically $2\pi$.
The sum of the baryonic charges of the corner around one intersection point is
the mean value of the baryonic charges of the faces surrounding that point.
In the case of Fig.~\ref{s_mean},
the sum is
\begin{equation}
\begin{split}
\frac{\theta}{2\pi}S_1+\frac{\pi-\theta}{2\pi}S_2+
\frac{\theta}{2\pi}S_3+\frac{\pi-\theta}{2\pi}S_4
&=\frac{\theta}{2\pi}(S_1+S_3)+\frac{\pi-\theta}{2\pi}(S_2+S_4)\\
&=\frac{\theta}{2\pi}(S_1+S_3)+\frac{\pi-\theta}{2\pi}(S_1+S_3)\\
&=\bar{S}_I.
\end{split}
\end{equation}
Hence $\sum S_A =\sum \bar{S}_I$ holds, 
and \eqref{sum_beta_general2} is zero.

Thus we have $d-1$ vanishing linear combinations of $\beta$ functions,
and this implies that there are $d-1$ RG invariant
parameters
\begin{equation}
f^{(I)}=\sum_aS^{(I)}_a\frac{1}{Ng^{'2}_a}
-\sum_kS^{(I)}_k\log h_k,
\label{rginvariant2}
\end{equation}
where $I=1,\ldots,d-1$ labels
linearly independent baryonic number assignments.
The coupling $g_a'$ is obtained as follows by integrating (\ref{gprime}):
\begin{equation}
\frac{1}{g^{'2}_a}=\frac{1}{g_a^2}-\frac{N}{8\pi^2}\log\left(\frac{1}{g_a^2}\right).
\end{equation}

We now comment on two special number assignments.
The first one is
\begin{equation}
S^{(1)}_A=1
\quad\forall A,\quad
b^{(1)}_\mu=0
\quad\forall \mu.
\label{allsone}
\end{equation}
This gives the RG invariant coupling
\begin{equation}
f^{(1)}=\sum_a\frac{1}{Ng^{'2}_a}
-\sum_k\log h_k.
\label{diag}
\end{equation}
Roughly speaking, this is related to the gauge coupling $g_{\rm diag}$
of the diagonal $SU(N)$ subgroup by $f^{(1)}\sim 1/(Ng_{\rm diag}^2)$.
The second one is
\begin{equation}
S^{(2)}_A=Q\quad
\mbox{for $(N,Q)$ face $A$},\quad
b^{(2)}_\mu=1\quad\forall\mu,
\label{allbone}
\end{equation}
which gives the RG invariant parameter
\begin{equation}\label{f2}
f^{(2)}=\sum_k\pm \log h_k.
\end{equation}
Because this depends only on the superpotential couplings
and does not include gauge couplings,
we can regard this
as a generalization of the $\beta$ deformation.\cite{Leigh:1995ep}

In ${\cal N}=1$ gauge theory, all couplings belong to
chiral multiplets and are complex.
The superpotential couplings are
complex, and the gauge couplings are also combined
with the $\theta$-angles to form the complex
couplings
\begin{equation}
\tau_a=\frac{\theta_a}{2\pi}+\frac{4\pi i}{g_a^2}.
\end{equation}
Thus, we have in total $d-1$ complex exactly marginal deformations.

%%%%%%%%%%%%%%%%%%%%%%%%%%%%%%%%%%%%%%%%%%%%%%%%%%%%%%%
\section{Brane configurations}\label{brane.sec}
In this section, we investigate the moduli parameters of brane configurations
with the help of BPS conditions.
The shape of the brane system depends on the ratio
of the tension of D5-branes to that of NS5-branes,
which is simply the string coupling constant $g_{\rm str}$.
In the construction of gauge theories, we take the decoupling limit
of gravity, which sends $g_{\rm str}$ to zero.
We first study the brane configuration in this limit in
\S\S\ref{weak.sec}, \ref{bps.sec} and \ref{sec.gc}.
Then we determine the number of moduli parameters
of brane systems
for general tilings.
In some simple cases, we explicitly solve the BPS conditions 
and determine the constraints imposed on the coefficients of
the equation defining the NS5-brane worldvolume.
It is, however, difficult to solve the BPS conditions in general cases.
This difficulty can be avoided by considering
the opposite limit, $g_{\rm str}\rightarrow\infty$,
which we discuss in \S\ref{strong.sec}.
In this limit, we can easily solve the BPS conditions,
and we obtain results similar to those in the weak coupling case.

%%%%%%%%%%%%%%%%%%%%%%%%%%%%%%%%%%%%%%%%%%%%%%%%%%%%%%%%%%%%%%%
\subsection{Holomorphic curves and tilings}\label{weak.sec}
In the explanation given in the introduction,
we regarded the brane configuration as a stack of D5-branes
with NS5-branes attached.
This is appropriate for analysis in the large $g_{\rm str}$ limit,
in which the D5-brane tension is much greater than the NS5-brane tension,
and the attached NS5-branes deform the D5-branes only very little.
When we consider the weak coupling limit $g_{\rm str}\rightarrow 0$,
however,
it is more natural to regard the system
as an NS5-brane with D5-branes attached to it.
Let $\Sigma$ be the NS5-brane worldvolume,
or, more precisely,
the union of NS5 and $(\pm N,1)$-brane worldvolumes.
In the weak coupling limit, the D5-brane tension
becomes negligible compared to the NS5-brane tension,
and the worldvolume $\Sigma$ becomes a smooth two-dimensional
surface in $4567$ space.
This surface has no boundaries, except the punctures at infinity.

For the purpose of describing the surface $\Sigma$,
we define the complex coordinates
\begin{equation}
s=x^4+ix^5,\quad
t=x^6+ix^7,
\end{equation}
and also
\begin{equation}
x=e^{2\pi s},\quad y=e^{2\pi t}.
\end{equation}
For simplicity, we assume that the torus is a unit square 
and the periods of $x^5$ and $x^7$ are $1$. 
We also assume that the axion field vanishes.
It is well known that $\Sigma$ is given by
\begin{equation}
P(x,y)=0,
\label{holocurve}
\end{equation}
with
the Newton polynomial $P(x,y)$ associated with the toric diagram.
Given a toric diagram $D$
associated with the dual Calabi-Yau geometry,
the corresponding Newton polynomial is defined by
\begin{equation}
P(x,y)=\sum_{(k,l)\in D}c_{k,l}x^ky^l.
\label{newton}
\end{equation}
The surface $\Sigma$ is punctured, and the number of punctures is equal to 
the number of edges 
on the perimeter of $D$ [the number of legs in the $(p,q)$ web of $D$].
The genus of $\Sigma$ is equal to the number of internal lattice points
of $D$.
The shape of $\Sigma$
depends on the coefficients $c_{k,l}$ assigned to the points $(k,l)$ in the
diagram.
We can easily show that an NS5-brane wrapped on the
holomorphic curve
(\ref{holocurve}) preserves
${\cal N}=2$
supersymmetry.

In the weak coupling limit,
the brane tilings constructed on a flat tori
should not be identified with the real shapes of 
worldvolume of branes.
Rather we should regard them as ``projections'' of the
NS5-brane worldvolumes $\Sigma$ onto the torus.
These projections are termed algae%
\footnote{Actually, some people,
including Passare and Tsikh, previously considered this object
and called it ``coamoeba''} in Ref.~\citen{Feng:2005gw},
and their ``tropical-like'' reduction gives a brane tiling.
The projected shape of the NS5-brane onto the D5-brane can be reproduced
from bipartite graphs by the method of Ref.~\citen{Hanany:2005ss}.
For our purpose, it is convenient to define
a function $Q(x^5,x^7)$ on the torus
as follows.
Let $\Pi(x^5,x^7)$ be the two-dimensional plane along the $46$ directions
with fixed $x^5$ and $x^7$ coordinates
in the $4567$ space.
In a four-dimensional space, two two-dimensional surfaces generically
intersect at isolated points, and we can thus define the intersection number of them.
We define the function $Q(x^5,x^7)$ as the intersection number
of the holomorphic curve $\Sigma$ and the plane $\Pi(x^5,x^7)$.
By definition, this function takes integer values at generic points
and jumps by one along cycles which are
defined as the projections of boundaries (punctures) of $\Sigma$
at infinity.
We can use the function $Q(x^5,x^7)$ to describe
the asymptotic structure of the surface $\Sigma$.
If the function $Q(x^5,x^7)$ takes only the values $0$ and $\pm1$,
we can construct the corresponding bipartite graph in the manner mentioned in the introduction.
But, in general, $Q(x^5,x^7)$ may not satisfy this condition.
Because $Q(x^5,x^7)$ is defined as a projection,
the condition $|Q(x^5,x^7)|\geq2$ does not imply the emergence of $(N,k)$-branes with $|k|\geq2$.
In the weak coupling limit, the worldvolume of an NS5-brane is a holomorphic curve,
and such branes never appear.

To clarify the relation between the structure of the surface $\Sigma$
and the function $Q(x^5,x^7)$,
let us focus on the asymptotic form of $\Sigma$.
Let us label the edges of the toric diagram
by $\mu=1,2,\ldots,d$ as we go around the sides in a counterclockwise manner.
We also define the primitive integral normal vector
$(m_\mu,n_\mu)$ of the $\mu$-th edge.
For each edge $\mu$, we have an external line
with direction $(m_\mu,n_\mu)$ in the web-diagram
constructed on the $46$ plane.
If a side $A$ of a toric diagram consists of $n_A$ edges,
we have $n_A$ parallel external lines associated with this side.
The corresponding asymptotic parts of $\Sigma$ are given by
\begin{equation}
P_A(x,y)=0,
\label{pazero}
\end{equation}
where $P_A(x,y)$ is the ``restriction'' of
the Newton polynomial to the side $A$
defined by
\begin{equation}
P_A(x,y)=\sum_{(k,l)\in A}c_{k,l}x^ky^l.
\label{newtona}
\end{equation}
Here, the summation is taken over only the points
on the side $A$.
Thus, the parameters responsible for the asymptotic behavior
of $\Sigma$ are those assigned to the points on the
perimeter of the toric diagram.
These variables
are actually redundant.
For example, we can use the freedom of rescaling $x$ and $y$
to eliminate two parameters.
Moreover, one coefficient can be eliminated through an overall
rescaling of $P(x,y)$.
Thus, we have $d-3$ complex parameters
associated with the asymptotic form of $\Sigma$.

Instead of the coefficients in the Newton polynomial,
it is more convenient to introduce another set of
variables which is directly related to the
asymptotic structure of the surface $\Sigma$.
The surface $\Sigma$ has $d$ punctures
corresponding to each external line in the web-diagram.
The asymptotic form of $\Sigma$ around the puncture $\mu$
is a cylinder.
Its projection onto the $4$-$6$ plane
gives an external line of the web-diagram,
and its projection onto the $5$-$7$ plane
gives a cycle $\alpha_\mu$ in the torus.
We define the parameters $M_\mu$ and $\zeta_\mu$,
representing the positions of the lines and the cycles, respectively, by
\begin{equation}
M_\mu=\lim_{\rightarrow\mu}(n_\mu x^4-m_\mu x^6),\quad
\zeta_\mu=\lim_{\rightarrow\mu}\left(n_\mu x^5-m_\mu x^7+\frac{1}{2}\right),
\label{mzeta2}
\end{equation}
or equivalently,
\begin{equation}
-e^{2\pi(M_\mu+i\zeta_\mu)}=\lim_{\rightarrow\mu}
x^{n_\mu}y^{-m_\mu}.\label{mzeta}
\end{equation}
These are defined as the limits in which we approach the puncture $\mu$
in the curve $\Sigma$.
At this point, only the fractional part of $\zeta_\mu$ is defined,
because the $x^5$ and $x^7$ coordinates are periodic.
The parameter $M_\mu$ determines the position in the $4$-$6$-plane
of the external line of the web-diagram corresponding to
a puncture $\mu$.
This is also regarded as the `moment' generated by the
tension of the $\mu$-th external line.
The parameter $\zeta_\mu$ determines the position of
the cycle $\alpha_\mu$ in the torus.
We have a set of $2d$ real parameters $\{M_\mu,\zeta_\mu\}$.
These parameters must be related to the $d-3$ complex parameters
in the Newton polynomial $P(x,y)$ associated with the perimeter
of the toric diagram.
Due to the translational symmetry, only $2d-4$ parameters
in $\{M_\mu,\zeta_\mu\}$
are relevant to the shape of the surface $\Sigma$.
We still have two more real parameters in $\{M_\mu,\zeta_\mu\}$.

The discrepancy described above is resolved if we take account of
constraints imposed on $\{M_\mu,\zeta_\mu\}$.
As mentioned above, $M_\mu$ can be regarded as
the moments acting on the brane system.
Because the brane configurations
we are considering here are BPS and stable,
the total moment must vanishes:
\begin{equation}
\sum_{\mu=1}^d M_\mu=0.
\label{momentzero}
\end{equation}
This can be directly proved as follows.
Because $\Sigma$ is holomorphic,
the pull-back of the $(2,0)$-form $ds\wedge dt$ onto
$\Sigma$ vanishes:
\begin{equation}
ds\wedge dt|_\Sigma=0.
\end{equation}
We can decompose this into the following two equations:
\begin{eqnarray}
(dx^4\wedge dx^7-dx^6\wedge dx^5)|_\Sigma&=&0,\label{4765}\\
(dx^4\wedge dx^6-dx^5\wedge dx^7)|_\Sigma&=&0.\label{4657}
\end{eqnarray}
With Stokes' theorem,
we can rewrite the integration of (\ref{4765}) over $\Sigma$
as
\begin{equation}
0
=\int_\Sigma(dx^4\wedge dx^7-dx^6\wedge dx^5)
=\int_{\partial\Sigma}(x^4dx^7-x^6dx^5)
=\sum_{\mu=1}^d \lim_{\rightarrow\mu} (n_\mu x^4-m_\mu x^6),
\label{momentcancel}
\end{equation}
and this is identical to the relation (\ref{momentzero}).
In the final step
of (\ref{momentcancel}),
we have used the fact that
the boundary of the surface $\Sigma$ is the union of
cycles $\alpha_\mu$,
and the integral of $(dx^5,dx^7)$ over a cycle $\alpha_\mu$
gives the integral vector $(m_\mu,n_\mu)$.

We also obtain a similar constraint on the parameters $\zeta_\mu$.
Using the relation (\ref{4657}), we can show
\begin{equation}
\int_{{\mathbb T}^2} dx^5dx^7Q(x^5,x^7)
=\int_\Sigma dx^5\wedge dx^7
=\int_\Sigma dx^4\wedge dx^6
=\int_{\partial\Sigma}x^4dx^6=0.
\label{intqcond}
\end{equation}
In other words, the average of the function $Q(x^5,x^7)$
on the torus vanishes.
In the final step, we have used the fact that the
punctures of $\Sigma$ are points in the $x^4$-$x^6$ plane,
and the integral vanishes.
Because the function $Q(x^5,x^7)$ is defined as
a step function that is discontinuous along
the cycles $\alpha_\mu$, whose positions are
determined by the parameters $\zeta_\mu$,
(\ref{intqcond}) imposes one constraint on the
parameters $\zeta_\mu$.

As mentioned above, the function $Q(x^5,x^7)$
is a step function which jumps by one
along the cycles $\alpha_\mu$.
We can decompose this function into step functions $q_\mu(x^5,x^7)$
associated with each cycle $\alpha_\mu$;
\begin{equation}
Q(x^5,x^7)=\sum_{\mu=1}^d q_\mu(x^5,x^7).
\label{qbyq}
\end{equation}
The function $q_\mu(x^5,x^7)$
jumps by one on the cycle $\alpha_\mu$,
whose position on the torus is specified by the
parameter $\zeta_\mu$.
The explicit form of the function $q_\mu(x^5,x^7)$ is
\begin{equation}
q_\mu(x^5,x^7)=[[-n_\mu x^5+m_\mu x^7+\zeta_\mu]],
\label{qdef}
\end{equation}
where $[[\cdots]]$ is defined by
\begin{equation}
-1/2\leq [[x]]-x\leq 1/2,\quad
[[x]]\in{\mathbb Z},
\end{equation}
for a real variable $x$.
The function $q_\mu(x^5,x^7)$ is not periodic but satisfies
\begin{equation}
q_\mu(x^5+p,x^7+q)=q_\mu(x^5,x^7)
+qm_\mu-pn_\mu,
\end{equation}
for an arbitrary integral vector $(p,q)$.
Note that the definition of $q_\mu(x^5,x^7)$ depends on
not only the fractional part of $\zeta_\mu$ but also on its
integral part.
In Eq.~(\ref{mzeta2}) we defined only the fractional part of
the parameters $\zeta_\mu$.
Let us define the parameter $\zeta_\mu$ including
the integral part by
\begin{equation}
\zeta_\mu=\int_{{\cal F}_0}q_\mu(x^5,x^7)dx^5dx^7,
\label{zteaq}
\end{equation}
where ${\cal F}_0$ is the specific fundamental region
${\cal F}_0=\{(x^5,x^7)|-1/2\leq x^5,x^7\leq 1/2\}$.
We can easily check that this definition gives
the same fractional part as the previous definition (\ref{mzeta2})
of the parameters.
The integral parts of $\zeta_\mu$ defined in this way
contribute to the function $Q(x^5,x^7)$ through
(\ref{qbyq}).

Now we can rewrite the constraint (\ref{intqcond})
imposed on the function $Q(x^5,x^7)$
in terms of the parameters $\zeta_\mu$.
Substituting the relation (\ref{qbyq}) into (\ref{intqcond}),
and using (\ref{zteaq}),
we obtain
\begin{equation}
\sum_{\mu=1}^d\zeta_\mu=0,
\label{t2cond}
\end{equation}
and this decreases the number of independent parameters
by one.
As a result, we have $d-3$ physical degrees of freedom
associated with the parameters $\zeta_\mu$.

The curve is specified by the parameters $M_\mu$
and $\zeta_\mu$ and the coefficients $c_{k,l}$ for the internal points
of the toric diagram.
In the next section,
we show that the BPS conditions of the D5-branes
fix some of them and leave $d-3$ parameters unfixed.
In \S\ref{sec.gc} we explicitly show in simple examples that
the free parameters are $\zeta_\mu$, subject to the
constraint (\ref{intqcond}).

%%%%%%%%%%%%%%%%%%%%%%%%%%%%%%%%%%%%%%%%%%%%%%%%%%%%%%
\subsection{BPS conditions}\label{bps.sec}
To this point, we have considered only the NS5-brane
in a system.
To realize the gauge theory,
we need to introduce D5-branes,
and these, in fact, impose extra constraints
on the parameters of the surface $\Sigma$.

As we mentioned above,
an NS5-brane wrapped on $\Sigma$ preserves the ${\cal N}=2$ supersymmetry.
Let $\epsilon_1$ and $\epsilon_2$ be the parameters of two supersymmetries.
If we introduce D5-branes into this system,
each of them breaks the supersymmetry down to ${\cal N}=1$,
which is specified by
$\epsilon_2=(Z/|Z|)\epsilon_1$,
where $Z$ is the central charge of the D5-brane,
given by
\begin{equation}
Z=\int_{\rm D5} ds\wedge dt
 =\oint_{\partial\rm D5} \log x\frac{dy}{y}.
\label{zintdsdt}
\end{equation}
Note that the topology of each D5-brane is a disk,
and its boundary $\partial\rm D5$ is a one cycle on the NS5-brane $\Sigma$.

In the context of the mirror Calabi-Yau
geometry,
this integral is simply the period.
By taking the T-duality
along one of the directions transverse to the brane system,
say the $x^8$ direction,
the NS5-brane is mapped to the Calabi-Yau $P(x,y)=uv$,
where $u,v \in \mathbb{C}$, and we have a new $\mathbb{S}^1$-fiber
from T-duality.
For the Calabi-Yau manifold, we have a holomorphic 3-cycle $\Omega$. In our coordinates, $\Omega$ can be written as
\begin{equation}
\Omega=\frac{dx}{x}\wedge\frac{dy}{y}\wedge\frac{du}{u}.
\end{equation}
We can trivially integrate this $\Omega$ along the $uv$-fiber direction,
and we are left with
\begin{equation}
\int_{{\rm D6}}\Omega
=\int_{{\rm D6}} \frac{dx}{x}\wedge\frac{dy}{y}\wedge\frac{du}{u}
\propto \int_{\rm D5} \frac{dx}{x}\wedge\frac{dy}{y}
=\oint_{\partial {\rm D5}} \log(x) \frac{dy}{y}.
\end{equation}
This is the central charge in (\ref{zintdsdt}).%
\footnote{The argument that
the period reduces to the integral
of a differential on 1-cycle of torus appeared long ago in Ref.~\citen{KLMVW},
although in a slightly different context.}

As we see below, for complicated examples,
it is difficult to directly analyze the shape of the parameter space.
It is still possible, however, to count the number of dimensions of that space.
Let us define the following quantities:
$d$ is the number of lattice points on the boundary of a toric diagram;
$I$ is the number of lattice points inside a toric diagram;
and $S$ is the area of a toric diagram.
Since we have one complex coefficient for each term
of a Newton polynomial, i.e., for each lattice point of a toric diagram,
we have $I+d$ complex parameters.
These variables
are actually redundant.
For example, we can use the freedom of rescaling $x$ and $y$
to eliminate two parameters.
Moreover, one coefficient can be eliminated through the overall
rescaling of $P(x,y)$.
Thus we have $I+d-3$ complex parameters,
or $2(I+d-3)$ real parameters.

Let $n_g$ be the number of $SU(N)$ gauge groups.
For the brane system to be BPS,
all the central charges $Z_i$ ($i=1,\ldots, n_g$) of the D5-branes must have the same argument:
\begin{equation}
\arg Z_1=\arg Z_2=\cdots =\arg Z_{n_g}.
\label{zreal}
\end{equation}
Since the number of D5-branes is twice the area of the toric diagram,
we have $2S-1$ conditions.
Thus we compute
\begin{equation}
2(I+d-3)-(2S-1)
=d-3,
\end{equation}
where we have used 
Pick's theorem,
\begin{equation}
S=I+\frac{d}{2}-1.
\end{equation}

In order to use (\ref{zintdsdt})
to compute the central charges $Z_i$ of the D5-branes,
we need to know the homotopy
classes of the D5-boundaries on the surface $\Sigma$.
There is a simple method, a
so-called {\it untwisting} \cite{Feng:2005gw}, 
to obtain them.

\begin{figure}[t]
\centerline{\includegraphics[scale=0.5]{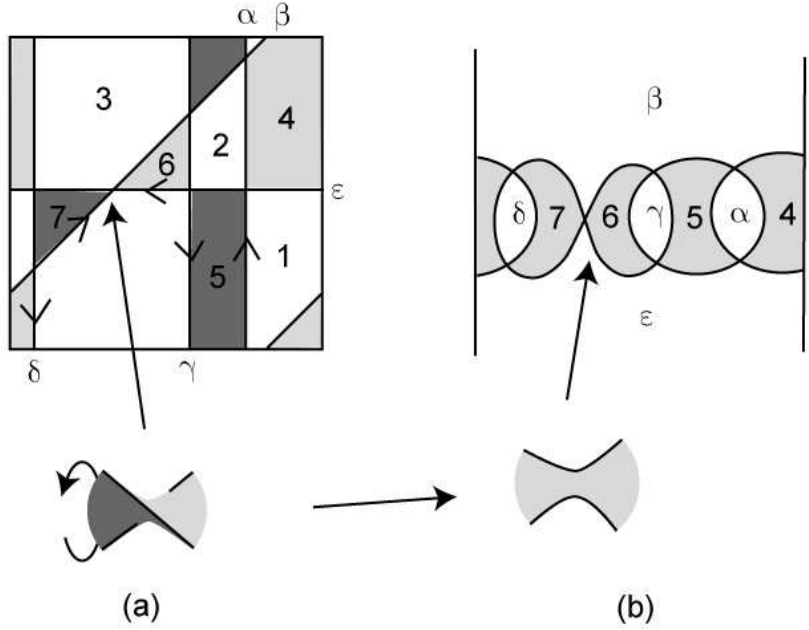}}
\caption{The untwisting operation of the SPP tiling.
The light and dark shaded faces in (a) represent $(N,-1)$ and $(N,+1)$
faces, respectively.
The result of the untwisting is shown in (b).
The two vertical lines on the left and right sides are identified.
Through the untwisting, $(N,-1)$ faces are turned over and become $(-N,1)$
faces. As a result, all shaded faces in (b) have NS5 charge $+1$.}
\label{untwistspp.eps}
\end{figure}

Because the untwisting is a topological operation,
we do not have to consider the real shapes of branes,
and we can regard a tiling constructed on a torus as
a real brane system.
Before the untwisting,
we remove the D5-branes, and we regard the white faces
in Fig.~\ref{untwistspp.eps}(a) as holes.
This makes the torus an NS5-brane composed of pieces connected at
the corners.
In Fig.~\ref{sppima.eps}(a), the pieces of
NS5-branes
are represented as shaded faces.
Because of the bipartite property of the system, any two shaded faces contacting each other
have opposite NS5 charges.
This means that the NS5-brane changes the orientation at the contact points
of faces.
In other words, the NS5-brane is ``twisted'' at the intersections of cycles.

In order to obtain the surface $\Sigma$, we ``untwist'' the surface
by turning the $(N,-1)$ faces over so that they
become $(-N,1)$ faces, as shown in Fig.~\ref{untwistspp.eps}(b).
By shrinking the holes to punctures, we finally obtain the surface $\Sigma$.
In the SPP case depicted in Fig.~\ref{untwistspp.eps},
we obtain a genus $0$ surface (i.e. a sphere) with the five
punctures corresponding to five cycles in the original tiling.

%%%%%%%%%%%%%%%%%%%%%%%%%%%%%%%%%%%%%%%%%%%%%%%%
\subsection{Example: generalized conifolds}\label{sec.gc}
As simple examples, we determine the parameter spaces
of brane configurations corresponding to generalized conifolds,
which include ${\mathbb C}^3$, the conifold, SPP, etc.,
as special cases.
First, we consider the SPP case, and then, we
discuss the generalization.

The Newton polynomial for SPP is
\begin{equation}
P(x,y)=y(x-x_\alpha)+(x-x_\gamma)(x-x_\delta).
\end{equation}
We have used the rescaling of $y$ and $P(x,y)$ to set the coefficients
of $xy$ and $x^2$ terms to $1$.
We can also set one of the quantities $x_\alpha$, $x_\gamma$, and $x_\delta$ to
an arbitrary value
through a rescaling of the variable $x$.
Here we set $x_\alpha=-1$.
The surface $\Sigma$ is a sphere with five punctures
[see Fig.~\ref{untwistspp.eps}(b)].
Let us use $x$ as a holomorphic coordinate of the Riemann sphere.
If we regard $x^4$ and $x^5$ as the latitude and the longitude
on the sphere,
we have two punctures, one at the north pole ($x=\infty$) and one at the south pole
($x=0$), which correspond to the cycles $\beta$ and $\epsilon$, respectively.
The three other punctures are at $x=x_\alpha=-1$, $x=x_\gamma$, and $x=x_\delta$.
The parameters $M_\mu$ and $\zeta_\mu$
are given by
\begin{eqnarray}
&&
e^{2\pi(M_\alpha+i\zeta_\alpha)}=
e^{2\pi(M_\beta+i\zeta_\beta)}=1,
\nonumber\\&&
e^{2\pi(M_\gamma+i\zeta_\gamma)}=-\frac{1}{x_\gamma},\quad
e^{2\pi(M_\delta+i\zeta_\delta)}=-\frac{1}{x_\delta},\quad
e^{2\pi(M_\epsilon+i\zeta_\epsilon)}=x_\gamma x_\delta.
\end{eqnarray}
We can easily see that the constraints (\ref{momentzero})
and (\ref{t2cond}) actually hold.

\begin{figure}[t]
\centerline{\includegraphics[scale=0.5]{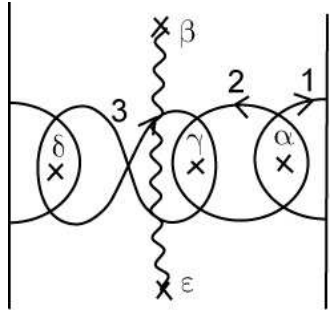}}
\caption{Three contours obtained from the boundaries
of the three $(N,0)$ faces, `1', `2', and `3',
in Fig.~\ref{untwistspp.eps}(a).
The symbols $\times$ indicate punctures corresponding to
the cycles in the tiling, and
the wavy line between the two punctures $\beta$ and $\epsilon$
represents a branch cut of $\log x$.
}
\label{sppcontour.eps}
\end{figure}

By tracing the boundaries of the three $(N,0)$ faces
[the white faces in Fig.~\ref{untwistspp.eps}(a)]
in the untwisting procedure,
we obtain three contours on the Riemann sphere
(see Fig.~\ref{sppcontour.eps}).
We should evaluate the integral (\ref{zintdsdt})
along these contours.
The function $\log x$ has a cut along a meridian
(the wavy line in Fig.~\ref{sppcontour.eps}),
and the differential
\begin{equation}
\frac{dy}{y}=
\frac{dx}{x-x_\gamma}
+\frac{dx}{x-x_\delta}
-\frac{dx}{x-x_\alpha}
\label{yrational}
\end{equation}
has poles at three punctures
$\alpha$, $\delta$, and $\gamma$.
\begin{figure}[b]
\centering{\includegraphics[scale=0.4]{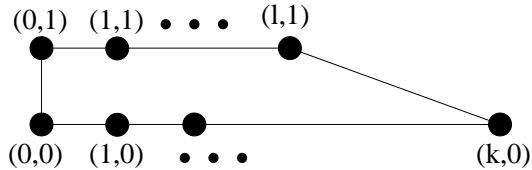}}
\caption{Toric diagram of a generalized conifold.}
\label{gconifoldTD}
\end{figure}
As shown in Fig.~\ref{sppcontour.eps},
each contour goes around two of these three punctures, 
and the integrals along them pick up the
residues of the differential $dy/y$.
We easily obtain
\begin{subequations}
\begin{eqnarray}
Z_1=\oint_1\log x\frac{dy}{y}
&=&2\pi i(\Log x_\alpha-\Log x_\delta),\\
Z_2=\oint_2\log x\frac{dy}{y}
&=&2\pi i(\Log x_\gamma-\Log x_\alpha),\\
Z_3=\oint_3\log x\frac{dy}{y}
&=&2\pi i(\Log x_\delta-\Log x_\gamma+2\pi i),
\end{eqnarray}
\end{subequations}
where $\Log$ is the principal value of the logarithm
defined with the branch cut in Fig.~\ref{sppcontour.eps}.
We have the extra $2\pi i$ in the third equation because
the contour crosses the cut.
For these central charges to have the same arguments,
the three punctures $\alpha$, $\gamma$ and $\delta$
must be at the same latitude:
\begin{equation}
1=|x_\alpha|=|x_\delta|=|x_\gamma|.
\end{equation}
(Note that we set $x_\alpha=-1$.)
This implies that all the parameters $M_\mu$ vanish,
and the unfixed parameters are
$\zeta_\gamma$ and $\zeta_\delta$.

We can easily generalize this analysis to the case of a generalized conifold
$x^ky^l=uv$ with arbitrary $k$ and $l$.
The toric diagram of a generalized conifold 
is displayed in Fig.~\ref{gconifoldTD}. 
The toric diagram has $k$ edges on the bottom and $l$ edges on the top,
and we denote the corresponding cycles by $\alpha_i$ and $\beta_i$, respectively.
We also have two cycles, $\gamma$ and $\delta$,
corresponding to the other two edges of the toric diagram.
The Newton polynomial is then
\begin{equation}
P(x,y)
=\prod_{i=1}^k(x-x_{\alpha_i})+y\prod_{i=1}^l(x-x_{\beta_i})=0,
\end{equation}
where $x_{\alpha_i}$ and $x_{\beta_i}$ are the positions of the punctures $\alpha_i$ and $\beta_i$.
We set the coefficients of $x^k$ and $yx^l$ terms to $1$
by rescaling $y$ and $P(x,y)$.
We can also set either $x_{\alpha_i}$ or $x_{\beta_i}$ to an arbitrary value.
We here set $x_{\alpha_1}=-1$.
The normal vectors $(m_\mu,n_\mu)$ are given for each cycle by
\begin{equation}
\alpha_i:(0,-1),\quad
\beta_i:(0,1),\quad
\gamma:(-1,0),\quad
\delta:(1,k-l).
\end{equation}
The parameters $M_\mu$ and $\zeta_\mu$ are given by
\begin{eqnarray}
&&
e^{2\pi(M_{\alpha_i}+i\zeta_{\alpha_i})}=-\frac{1}{x_{\alpha_i}},\quad
e^{2\pi(M_{\beta_i}+i\zeta_{\beta_i})}=-x_{\beta_i},\nonumber\\
&&
e^{2\pi(M_\gamma+i\zeta_\gamma)}=\frac{\prod_{i=1}^k(-x_{\alpha_i})}{\prod_{i=1}^l(-x_{\beta_i})},\quad
e^{2\pi(M_\delta+i\zeta_\delta)}=1.
\end{eqnarray}
We use $x$ as a coordinate on the Riemann sphere $\Sigma$.
The two punctures $\delta$ and $\gamma$ are at the north ($x=\infty$) and south poles ($x=0$),
and the function $\log x$ has a cut along a meridian connecting them.
The other $k+l$ punctures are poles of
the differential
\begin{equation}
\frac{dy}{y}
=\sum_{i=1}^k\frac{dx}{x-x_{\alpha_i}}
-\sum_{i=1}^l\frac{dx}{x-x_{\beta_i}}.
\end{equation}

The tiling for the generalized conifold has $k$ down-going cycles $\alpha_i$ and $l$ up-going cycles $\beta_i$.
Before performing the untwisting operation,
we need to fix the order of these vertical cycles in the tiling.
The choice of the ordering corresponds to the choice of
the toric phase.
Let $\eta_i$ ($i=1,\ldots,k+l$) be the set of two
kinds of cycles $\alpha_i$ ($i=1,\ldots,k$) and
$\beta_i$ ($i=1,\ldots,l$) ordered as
\begin{equation}
\arg x_{\eta_1}\leq \arg x_{\eta_2}\leq\cdots\leq\arg x_{\eta_{k+l}}.
\end{equation}
The cycles $\eta_i$ divide the tiling into $k+l$ strips.
\begin{figure}[htb]
\centerline{\includegraphics[scale=0.5]{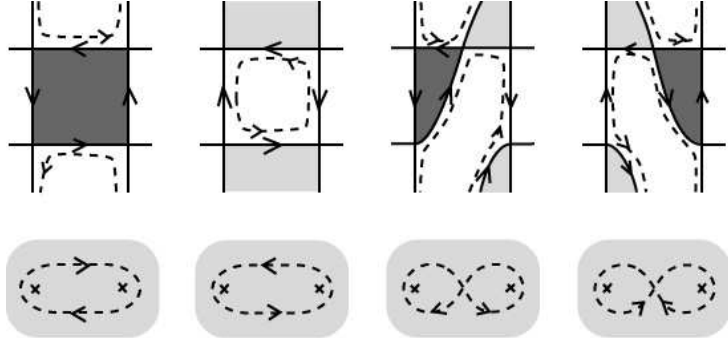}}
\caption{The upper four diagrams depict pieces of tilings of generalized conifolds.
The tiling of the generalized conifold $x^ky^k=uv$ consists of
$k+l$ of them.
Each strip has one $(N,0)$ face.
The dashed lines represent the boundaries of $(N,0)$ faces.
The lower four diagrams depict the contours obtained through the untwisting
operation from the boundaries.
Each of them encloses two punctures (indicated by $\times$)
corresponding to the two sides of the strip.
}
\label{gc.eps}
\end{figure}
Let us focus on the strip between the cycles $\eta_i$ and $\eta_{i+1}$.
There are four different types of strips, depending on the orientation of
the two sides of the strips (Fig.~\ref{gc.eps}).
In any case, the strip includes one $(N,0)$ face,
and the untwisting operation maps its boundary to
a contour on $\Sigma$ enclosing two punctures, $\eta_i$ and $\eta_{i+1}$
(see Fig.~\ref{gc.eps}).
We can easily show that the contour integral is given by
\begin{equation}
Z_i=\oint_{\eta_i}\log x\frac{dy}{y}=
\left\{\begin{array}{l}
2\pi i(\Log x_{\eta_{i+1}}-\Log x_{\eta_i})\quad\mbox{or}\\
2\pi i(\Log x_{\eta_{i+1}}-\Log x_{\eta_i}+2\pi i),
\end{array}\right.
\end{equation}
where the additional term $2\pi i$ is included when the contour crosses
the branch cut of $\log x$.
For the $k+l$ central charges $Z_i$ to satisfy the BPS conditions
(\ref{zreal}),
the following equations must hold:
\begin{equation}
1=|x_{\alpha_1}|=|x_{\alpha_2}|=\cdots=|x_{\alpha_k}|
=|x_{\beta_1}|=|x_{\beta_2}|=\cdots=|x_{\beta_l}|.
\end{equation}
(Note that we set $x_{\alpha_1}=-1$.)
This relation means that all the parameters $M_\mu$ vanish.
Therefore, the unfixed parameters are
$\zeta_{\alpha_i}$ ($i=2,\ldots,k$)
and $\zeta_{\beta_i}$ ($i=1,\ldots,l$),
which describe the positions of the poles $\alpha_i$ and $\beta_i$
aligned on the equator.
Because we first fixed the order of these poles,
the moduli space is a part of $\mathbb T^{d-3}$.
Different orders of the poles on the equator can be interpreted
as different toric phases related by the Seiberg duality.
The total moduli space defined as the union of all the phases
is $\mathbb T^{d-3}$.

The moduli space we obtain here is identical to
that obtained in the dual D4-NS5 system (the elliptic model)
in Ref.~\citen{Uranga:1998vf}.
\begin{table}[b]
\caption{Brane configuration in the elliptic model.
In that model, the 9-direction is compactified.}
\label{ellipticBC}
\begin{center}
\begin{tabular}{ccccccccccc}
\hline\hline
&0&1&2&3&4&5&6&7&8&9\\
\hline
D4 & $\circ$ & $\circ$ & $\circ$ & $\circ$ & & & & & &$[\circ$]\\
NS5$_a$ & $\circ$ & $\circ$ & $\circ$ & $\circ$ &$\circ$& $\circ$ & & & & \\
NS5$_b$ & $\circ$ & $\circ$ & $\circ$ & $\circ$ & &  &$\circ$&$\circ$&  & \\
\hline
\end{tabular}
\end{center}
\end{table}
The elliptic model is a brane configuration with D4-branes and NS5-branes in type IIA theory.
The brane configuration is given in Table \ref{ellipticBC}.
The direction $x_9$ is compactified on $\mathbb S^1$, and
$N$ D4-branes are wrapped on this $\mathbb S^1$.
There are two kinds of NS5-branes stretched in different directions, $(x^4,x^5)$ and $(x^6,x^7)$,
which we call NS5$_a$- and NS5$_b$-branes, respectively.
These NS5-branes intersect D4-branes at different points on $\mathbb S^1$
and divide the D4-branes into segments.
All branes stretch along the $(x^0,x^1,x^2,x^3)$ directions.
An $SU(N)$ gauge field exists on each segment of D4-branes separated by NS5-branes.
There are two bi-fundamental fields
at each intersection of D4-branes and NS5-branes.
If a segment of D4-branes ends on the same kind of NS5-branes,
this segment can move 
along the direction of these NS5-branes.
This is interpreted as an $SU(N)$ adjoint scalar field.
If the numbers of NS5$_a$-branes and NS5$_b$-branes are $k$ and $l$, respectively,
this gives the gauge theory described by the brane tiling obtained
from the toric diagram displayed in Fig.~\ref{gconifoldTD}.
Actually, we can obtain the generalized conifold
using T-duality in the direction $x^9$.

The moduli parameters of the elliptic model are
the $x^9$ coordinates of NS5$_a$-branes and NS5$_b$-branes,
and we can identify them as
$\arg x_{\alpha_i}$ and $\arg x_{\beta_i}$.
We can fix the position of one brane by translation,
and the number of degrees of freedom is $k+l-1=d-3$.
The moduli space is $\mathbb T^{d-3}$, parametrized by the relative positions of
$k+l$ points on $\mathbb S^1$, and
this is precisely the same as the moduli space
of the D5-NS5 system obtained above.

In the examples we have discussed in this subsection,
we have two kinds of parameters, $M_\mu$ and $\zeta_\mu$,
and we found that the BPS condition (\ref{zreal}) for D5-branes imposes conditions
on $M_\mu$, and we are left with $d-3$ free parameters among the $\zeta_\mu$.
How can this be generalized to general tilings?
In general, we have complex coefficients for internal points
of toric diagrams, in addition to $M_\mu$ and $\zeta_\mu$.
Considering the number of parameters and constraints,
we conjecture that all these additional parameters are fixed
by the BPS conditions, and the number of free parameters
among the $\zeta_\mu$ is still $d-3$, as in the example discussed above.
Unfortunately, we have not yet proved this conjecture,
but its validity is supported
by the analysis in the strong coupling limit
given in the next subsection.

%%%%%%%%%%%%%%%%%%%%%%%%%%%%%%%%%%%%%%%%%%%%%%%
\subsection{Strong coupling limit}\label{strong.sec}
In this subsection, we determine the parameter spaces of brane configurations
in the large $g_{\rm str}$ limit.
The reason we study this limit even though the relation to gauge theories is not clear is that
the analysis of the parameter space in this case is quite simple.
We can still use the parameters $M_\mu$ and $\zeta_\mu$ defined in (\ref{mzeta2})
to describe the asymptotic forms of brane configurations,
and we find below that all the $M_\mu$ are fixed, and
BPS configurations are parametrized by only $\zeta_\mu$
with the condition (\ref{t2cond}) imposed.

In the strong coupling limit,
the tension of D5-branes is much greater than that of NS5-branes,
and the D5 worldvolume wrapped on the ${\mathbb T}^2$ is almost flat.
This means that the D5 worldvolume is a point in the $x^4$-$x^6$
plane, and
all the external lines (the projection of the NS5-branes onto the $46$-plane)
of the web diagram meet at this point.
Therefore, we have no degrees of freedom to deform the web diagram.
This implies that $M_\mu=0$.
The parameters of the brane configuration are only
the positions of the NS5-branes in the internal space.
In other words, a brane configuration is
completely determined by $\zeta_\mu$, the
positions of the cycles in the tiling.
We investigate the constraint imposed on these parameters
by requiring that the brane preserve the supersymmetry.

We define the complex coordinate $z=x^5+ix^7$
to parametrize the worldvolume of D5-branes.
The shape of a brane is described by two real functions $f(z)$ and $g(z)$
as
\begin{equation}
x^4=f(z),\quad
x^6=g(z).
\end{equation}
Because the tension of D5-branes is so large,
we can assume that the deformation of the branes is small,
and here we omit higher-order terms of $f$ and $g$ in what follows.

We define the function $Q(z)$ which gives the NS5 charge
on the $z$-plane.
This is nothing but the function given in (\ref{qbyq}).
In \S\ref{weak.sec}, this function is defined as the ``projection''
of NS5-branes, while here it represents the NS5 charge of the worldvolume.
When $g_{\rm str}$ is very large,
the BPS condition is given by 
\begin{equation}
t_1^at_2^b(i\sigma_{ab})
=\varphi\sigma_z+\sigma_x,
\label{strongbps}
\end{equation}
where $\sigma_{ab}$ ($a=4,5,6,7$) are the chiral parts of
the antisymmetric products of the gamma matrices $\gamma^{ab}$ in 4567 space.
[See the appendix for a derivation of (\ref{strongbps}) and conventions.]
Here, $\varphi$ is the angle defined by
\begin{equation}
\varphi=\arg\left(N+\frac{iQ(z)}{g_{\rm str}}\right)\sim\frac{1}{Ng_{\rm str}}Q(z),
\label{strongphi}
\end{equation}
and $t_1^a$ and $t_2^a$ are the following tangent vectors
of the worldvolume:
\begin{equation}
t_1^a=(\partial_{x^5}f,1,\partial_{x^5} g,0),
\quad
t_2^a=(\partial_{x^7}f,0,\partial_{x^7}g,1).
\label{t1t2}
\end{equation}
Here we assume that the axion field vanishes.
Substituting
(\ref{t1t2}) into the BPS condition
(\ref{strongbps}), we obtain
\begin{equation}
\sigma_{57}
+(\partial_{x^5}f)\sigma_{47}+(\partial_{x^5}g)\sigma_{67}
+(\partial_{x^7}f)\sigma_{54}+(\partial_{x^7}g)\sigma_{56}
=\varphi\sigma_z+\sigma_x.
\end{equation}
Then, using the explicit form of $\sigma_{ab}$ given in the appendix
and introducing 
the complex function $F=2Ng_{\rm str}(g+if)$,
the above conditions become
\begin{equation}
\frac{\partial F}{\partial z^*}
=Ng_{\rm str}\varphi
=Q(z).
\label{qnscons}
\end{equation}

Let us integrate (\ref{qnscons}) over a fundamental region ${\cal F}$
of the $z$-plane.
The integral of the left-hand side can be rewritten as a contour
integral using Stokes' theorem:
\begin{equation}
\int_{\cal F}\frac{\partial F}{\partial z^*}d^2z
=\int_{\cal F}dz\wedge dF
=\int_{\partial\cal F}dz F.
\label{contint}
\end{equation}
In the final expression, the contribution of
opposite sides of the fundamental region cancel
and (\ref{contint}) always vanishes.
Therefore, 
for the existence of a solution of (\ref{qnscons}),
the function $Q(z)$ must satisfy
\begin{eqnarray}\label{ch-avr}
\int_{\cal F} Q(z)d^2z=0.
\end{eqnarray}
This is the same as (\ref{intqcond}).

In summary,
the brane configurations in the strong coupling limit are parametrized
by only the parameters $\zeta_\mu$,
with the condition (\ref{ch-avr}) imposed.
This is identical to the case of the weak coupling
limit.
We emphasize that in the analysis in the strong coupling limit
we considered general tilings.
Although we cannot explicitly determine the moduli space of general tilings
in the weak coupling limit, the result obtained here strongly suggests
that the moduli space is always parametrized by only the parameters $\zeta_\mu$,
and the other parameters, $M_\mu$ and $c_{k,l}$ associated with
the internal points of the toric diagrams are fixed by
the BPS conditions.

An important difference between the strong coupling and weak coupling limits is that
the function $Q(x^5,x^7)$ in the strong coupling limit is not the projection but
the actual NS5 charge.
Therefore we have $(N,k)$-branes with $|k|\geq2$
if $|Q(x^5,x^7)|\geq2$ at some points on the torus.
If such branes appear, we cannot use bipartite graphs to
determine the gauge groups and matter content.
In the following,
we solve the condition (\ref{ch-avr}) explicitly in two cases, 
the conifold and the SPP,
and we show that such tilings do appear in the latter example.

\begin{figure}[htb]
\centerline{\includegraphics[scale=0.5]{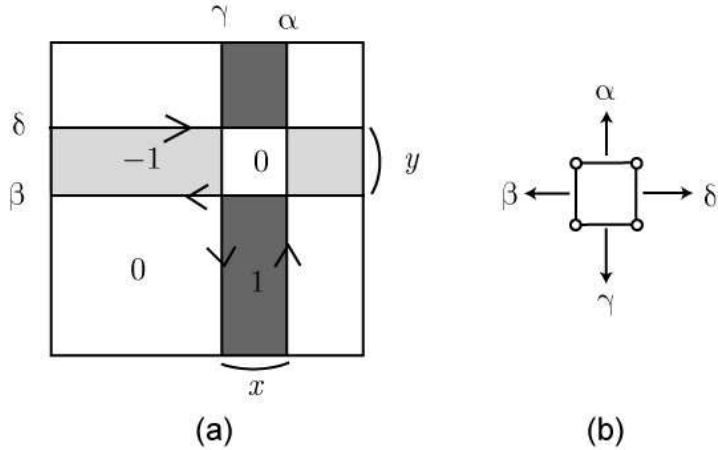}}
\caption{The tiling of the conifold (a) and the corresponding toric diagram (b).}
\label{conima.eps}
\end{figure}

First, we consider the conifold.
The toric diagram of the conifold and corresponding brane configuration
are given in Fig.~\ref{conima.eps}.
We can fix the positions of one horizontal and one vertical cycle
by using the translational symmetry,
and consequently there are only two physical parameters, $x$ and $y$,
satisfying $0\leq x,y\leq 1$,
which represent the separations of parallel cycles.
In order for the average charge to vanish,
we need both positive and negative charges,
and the unique charge assignment is given in Fig.~\ref{conima.eps}.
Furthermore,
(\ref{ch-avr}) demands that two shaded faces have the same area.
This is the case only when $x=y$.
Therefore, the parameter space is $\mathbb S^1$.
Note that this is consistent with our previous analysis in the weak coupling limit.
(The conifold represents the $k=l=1$ case of the generalized conifold in \S\ref{sec.gc}.)

\begin{figure}[htb]
\centerline{\includegraphics[scale=0.5]{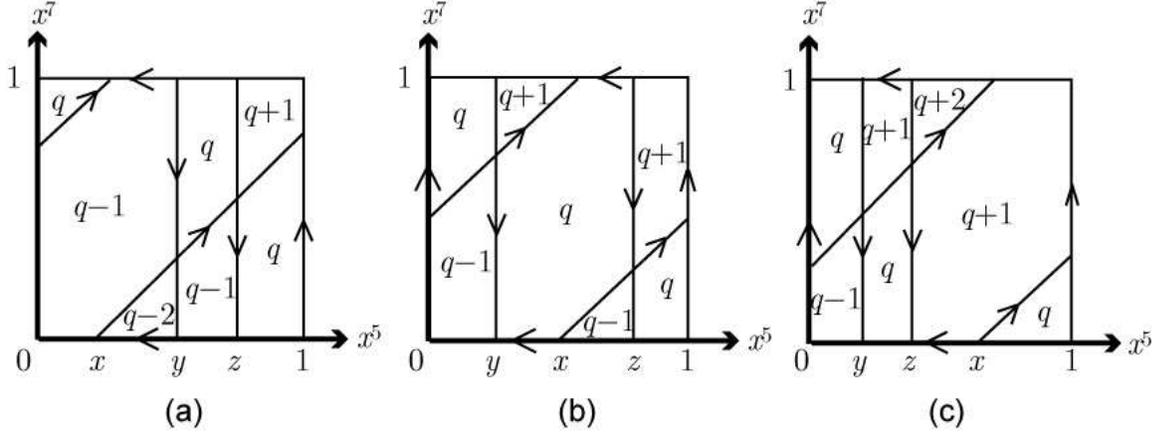}}
\caption{Three configurations for the brane tiling of the SPP case.}
\label{spp-iso.eps}
\end{figure}

\begin{figure}[t]
\centerline{\includegraphics[scale=.7]{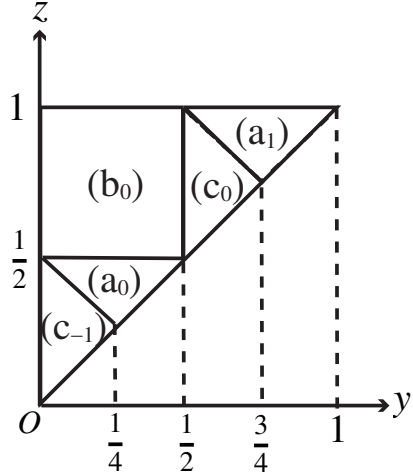}}
\caption{Phase diagram for the SPP case. 
(a$_q$), (b$_q$) and (c$_q$) refer to the configurations 
(a), (b) and (c), respectively, with charge $q$ 
in Fig.~\ref{spp-iso.eps}.}
\label{spp-phase-2.eps}
\end{figure}

The next example is the SPP, whose toric diagram is given in Fig.~\ref{sppima.eps}.
This toric diagram has 5 edges, and thus the number of physical parameters should be 2.
The corresponding tiling has 5 cycles.
We fix the positions of the vertical cycle $\alpha$ at $x^5=0$
and the horizontal cycle $\epsilon$ at $x^7=0$.
Let $x$, $y$ and $z$ be the $x^5$ coordinates of the intersections of the remaining cycles,
$\beta$, $\gamma$ and $\delta$, and the $x^5$ axis, respectively.
We choose $x$, $y$ and $z$ such that $0<x,y,z<1$ and $y<z$.
We can classify the configuration of tilings into three cases (See Fig.~\ref{spp-iso.eps}),
according to the order of $x$, $y$ and $z$.
By setting the charge on one face in the tiling, 
the charges on the other faces are automatically determined.
If we assign the charges as in Fig.~\ref{spp-iso.eps},
the BPS condition (\ref{ch-avr}) becomes
\begin{equation}\label{chvan-spp}
0=\int_{\cal F}Q(z)d^2z=\frac{1}{2}+q+x-y-z.
\end{equation}
Let us take $y$ and $z$ as independent variables.
Then, through the condition (\ref{chvan-spp}), 
we can uniquely determine $q$ and $x$ as functions of $y$ and $z$.
The result is depicted in the ``phase diagram'' appearing in Fig.~\ref{spp-phase-2.eps}.
The lines dividing the triangle in Fig.~\ref{spp-phase-2.eps}
represent the transition points
at which the parameter $x$ crosses $x=y$, $x=z$, and $x=0$
(or, $x=1$, which is identified with $x=0$).
The subscripts of the labels represent the charges $Q$.
In the region (b$_0$), the tiling has faces with NS5 charges $-1$, $0$ and $1$,
while in the regions (a$_q$) and (c$_q$), the tilings have faces with NS5 charge $+2$ or $-2$, in addition to $0$ and $\pm1$.
Such configurations cannot be translated into bipartite graphs.
Generally, in the large $g_{\rm str}$ limit,
such types of tilings are allowed.

%%%%%%%%%%%%%%%%%%%%%%%%%%%%%%%%%%%%%%%%%%%%%%%%%%%%%%%%%%%
\subsection{Wilson lines}
To this point, we have only discussed geometric deformations of
the brane configurations, and we have
obtained $d-1$ independent degrees of freedom.
These are the positions $\zeta_\mu$ $(\mu=1,\ldots,d)$
of $d$ cycles with one constraint imposed,
and two of them are unphysical, due to the translational invariance.
Because the brane configurations preserve
${\cal N}=1$ supersymmetry,
all scalar parameters must belong to chiral multiplets.
This implies that every real parameter
is combined with its `superpartner' to
form a complex parameter.

What is the partner of the parameter $\zeta_\mu$?
A natural guess is that it is the Wilson line associated with the cycle $\alpha_\mu$.
Because each cycle is the projection of the NS5-brane worldvolume wrapped on a $1$-cycle
on the torus, we can construct a gauge invariant quantity by
integrating the $U(1)$ gauge field $A$ on the NS5-brane along the cycle $\alpha_\mu$:
\begin{equation}
W_\mu=\oint_{\alpha_\mu} A.
\end{equation}
Actually, this guess is correct.
A direct way to prove this is to perform the supersymmetry transformation
of the fields on the NS5-brane.
The scalar, spinor and vector fields on an NS5-brane form a vector multiplet
of ${\cal N}=4$ supersymmetry, and the supersymmetry transformation for these fields
is a combination of the global supersymmetry in the bulk and the local $\kappa$-symmetry on the brane.
By determining how scalar fields enter into the supersymmetry transformation
of chiral spinor fields, we can determine the multiplet structure
of the bosonic fields on the brane.
Although such direct analysis is not difficult,
to avoid the trouble involved with treating spinors and gamma matrices,
here we adopt another approach, which uses S and T dualities to transform
the brane configuration to other ones, in which the multiplet structure
is clearer.

We have many NS5-branes with different orientations.
For concreteness,
we assume that the system consists of
D5-branes along 012357,
an NS5-brane along 012345,
and an NS5-brane along 012367,
and
we consider the multiplet structure of the bosonic fields
on the NS5-brane along 012345.
Let $A_\mu$ ($\mu=0,\ldots,5$) and $A_i$ ($i=6,7,8,9$)
be the vector and scalar fields on the NS5-brane, respectively.
We use the same symbol for both vector and scalar fields,
because we use T-duality transformations which mix
vector and scalar components.

Let us first focus on the two intersecting NS5-branes
along $012345$ and $012367$.
These two branes preserve ${\cal N}=2$ supersymmetry, and
the fields $A_M$ ($M=0,\ldots,9$) are divided into
a vector multiplet, $\{A_0,A_1,A_2,A_3,A_8,A_9\}$, and a hypermultiplet,
$\{A_4,A_5,A_6,A_7\}$.
This grouping is most clearly seen by performing
the $S$ and $T(4589)$ duality transformations.
[Note that $T(4589)$ is the T-duality transformation along the $4589$ directions.]
The S-duality transforms two NS5-branes into two D5-branes,
and the T-duality makes them space-filling D9-branes
and D5-branes along $012389$.
After the duality transformation, the Lorentz group $SO(1,9)$
is broken to $SO(1,5)\times SO(4)$, and the fields $A_M$
are divided correspondingly.

Next, let us consider another pair of branes, the NS5-brane along $012345$
and the D5-brane along $012357$.
We perform an S-duality transformation, followed by
$T(5)$.
This combination transforms the original NS5 and D5
into D4 along $01234$ and NS5 along $012357$.
This is the configuration used to describe
an ${\cal N}=2$ elliptic model.
(Precisely speaking, an extra compactification is needed to obtain the elliptic model.
This has no relation to the ${\cal N}=1$ elliptic models mentioned in
\S\ref{sec.gc}.)
If the D4-brane ends on the NS5-brane,
it can move along $5$ and $7$ directions,
and in the elliptic model, these are identified with the
scalar fields in an ${\cal N}=2$ vector multiplet.
This implies that the set $\{A_0,A_1,A_2,A_3,A_5,A_7\}$ forms a vector
multiplet, and the set $\{A_4,A_6,A_8,A_9\}$ belongs
to a hypermultiplet.

We obtained two decompositions for different choices of
the ${\cal N}=2$ supersymmetry.
We can obtain the multiplets of unbroken ${\cal N}=1$ supersymmetry
as the superposition of these two decompositions,
and we obtain a vector multiplet
$\{A_0,A_1,A_2,A_3\}$ and
three chiral multiplets
$\{A_4,A_6\}$, $\{A_5,A_7\}$, and $\{A_8,A_9\}$.
In particular, we find that the parameter $\zeta_\mu\sim A_7$ is
combined with the Wilson line $W_\mu\sim A_5$, as hypothesized above.

There is one constraint, (\ref{t2cond}), imposed on the parameters $\zeta_\mu$.
The multiplet structure requires that this also be the case for the Wilson lines
$W_\mu$.
Actually, with Stokes' theorem, we can show that
\begin{equation}
\sum_{\mu=1}^dW_\mu
=\sum_{\mu=1}^d\oint_{\alpha_\mu} A
=\int_\Sigma dA=0.
\end{equation}
In the final step, we have used $dA=0$.

We can also show that
two of the $d-1$ degrees of freedom of the Wilson lines
are redundant.
Let us consider gauge transformations
of the RR $2$-form field in the bulk.
They also change the gauge field on the NS5-brane as
\begin{equation}
\delta C_2=d\Lambda,\quad
\delta A=\Lambda.
\end{equation}
If the parameter $\Lambda$ is a closed $1$-form,
this transformation changes only $A$.
The Wilson lines are transformed as
\begin{equation}
W_\mu\rightarrow W_\mu'=W_\mu+\oint_{\alpha_\mu}\Lambda.
\label{wtow}
\end{equation}
We have two independent closed $1$-form on
the torus, and the gauge transformation
(\ref{wtow}) decreases the number of physical degrees of
freedom by two.

%%%%%%%%%%%%%%%%%%%%%%%%%%%%%%%%%%%%%%%%%%%%%%%%%%%%%%%%%
\section{Comparison of parameters}\label{comparison.sec}
In \S \ref{gauge.sec}, we analyzed the $\beta$-functions
of gauge and superpotential couplings, and
we found $d-1$ complex marginal deformations.
We also found that the corresponding brane configuration
has $d-3$ complex degrees of freedom
corresponding to changing the shape and Wilson lines.
In this section, we give an argument aimed at determining
the relations among these deformations and degrees of freedom.

To begin, let us consider how the couplings are obtained in string theory.
We first focus on the real parameters.
The gauge coupling for each $SU(N)_a$
can be read off of the Born-Infeld action
of D5-branes as
\begin{equation}
\frac{1}{Ng^{'2}_a}
=
\frac{1}{Ng_a^2}-\frac{1}{8\pi^2}\log\frac{1}{g_a^2}
\sim\frac{A'_a}{4\pi Ng_{\rm str}\alpha'}
-\frac{1}{8\pi^2}\log\frac{A'_a}{4\pi g_{\rm str}\alpha'}
,
\label{oong}
\end{equation}
where $A_a'$ is the area of the face $a$
evaluated with the real shape of the D5-brane worldvolume.
Then,
each term in the superpotential is
induced by the string worldsheet wrapped on the
corresponding $(N,\pm1)$ face and is roughly given by
\begin{equation}
-\log|h_k|
\sim
-\log\left(g_{\rm str}^{(n/2)-1}
e^{-A'_k/(2\pi\alpha')}\right)
=
\frac{A'_k}{2\pi\alpha'}
+\left(\frac{n}{2}-1\right)\log\frac{1}{g_{\rm str}},
\label{loghk}
\end{equation}
where $n$ is the number of fields included in the interaction
term.

Because of the difficulty involved in quantizing strings
in the background with D5- and NS5-branes,
we cannot obtain precise relations between
the parameters in gauge theories and
those in string theory.
The purpose of this section is to derive only rough
relations among them.
For this purpose,
we focus only on the first terms in the final expressions
in (\ref{oong}) and (\ref{loghk}).
Furthermore, we ignore the $N$ and $g_{\rm str}$ dependences
of these terms, and we simplify the relations
as
\begin{equation}
\frac{1}{N{g_a'}^2}\sim A_a,\quad
-\log|h_k|\sim A_k.
\label{coupling2area}
\end{equation}
In these relations,
in addition to the simplification
mentioned above,
we have replaced the areas $A_A'$,
which are evaluated with the real shapes of the D5-brane worldvolumes,
by the areas $A_A$ on a flat torus
divided by straight cycles.
This combination of simplification results in
an approximate treatment that is so rough
that we cannot obtain any quantitative results
from the analysis in the following.
On the other hand, these simplifications make the equations below quite simple
and clarify the qualitative relations among
the parameters.

Replacing the terms in
the RG invariant parameters (\ref{rginvariant2})
by the corresponding areas according to (\ref{coupling2area}),
we obtain
\begin{equation}
f^{(I)}=\sum_AS_A^{(I)}A_A
=\int_{\cal F} S^{(I)}(x^5,x^7)dx^5dx^7,
\label{saaa}
\end{equation}
where the function $S^{(I)}(x^5,x^7)$ is defined so that
$S^{(I)}(x^5,x^7)=S_A^{(I)}$ on the face $A$.
The integration region ${\cal F}$ is an arbitrary fundamental region.
Then, using the parameters $b_\mu$ in (\ref{ssb}) and
the functions $q_\mu(x^5,x^7)$ defined in (\ref{qdef}),
we obtain the function $S(x^5,x^7)$ as
\begin{equation}
S^{(I)}(x^5,x^7)=\sum_{\mu=1}^d b_\mu^{(I)} q_\mu(x^5,x^7)+c,
\end{equation}
where $c$ is a new parameter determining the constant part of $S^{(I)}$.
The periodicity of $S(x^5,x^7)$ is guaranteed by the condition
(\ref{balpha}).
If we substitute this into
(\ref{saaa}), we obtain
\begin{equation}
f^{(I)}=\sum_{\mu=1}^d b_\mu^{(I)}\zeta_\mu+c.
\label{sabz}
\end{equation}
This relation shows how the RG invariant parameters $f^{(I)}$
in the gauge theory
are related to the parameters $\zeta_\mu$ and $c$ in the
string theory.

Among the parameters $\zeta_\mu$ and $c$ in the string theory,
we know that $\zeta_\mu$ represent the positions of the cycles $\mu$
on the torus and describe the shape of the brane configuration.
What is the other parameter, $c$?
We can identify this degree of freedom
with the expectation value of the dilaton field in the following way.

If we substitute the assignment (\ref{allsone})
into (\ref{sabz}),
we obtain
\begin{equation}
c=f^{(1)}
\sim \frac{1}{Ng_{\rm diag}^2}.
\label{sabz1}
\end{equation}
Hence, we find that the parameter $c$ is the diagonal gauge coupling (\ref{diag}).
Roughly speaking, it
is the gauge coupling of the theory
on D5-branes wrapped on the torus without NS5-branes attached.
We can read off the coupling
from the action of the D5-branes, and we find
\begin{equation}
c=\frac{A_{\rm tot}}{\alpha'e^\phi}.
\end{equation}
With this equation, we can identify the parameter $c$
with the expectation value of $e^{-\phi}$.
(More precisely, $c$ depends not only on the dilaton but also
on the size of ${\mathbb T}^2$.)
This correspondence can easily be extended to the
correspondence between complex parameters.
The diagonal gauge coupling $1/g_{\rm diag}^2$ is combined with
the theta angle $\theta_{\rm diag}$ of the diagonal gauge group,
and we can read off the relation
$\theta_{\rm diag}\sim C_{57}$
from the D5-brane action.
We thus obtain the relation
\begin{equation}
\tau_{\rm diag}\sim ic\sim
C_{57}+\frac{i}{e^\phi}.
\end{equation}
We can show that
the right-hand side of this relation is actually the
scalar component of one chiral multiplet
by checking the transformation law of fermions
in type IIB supergravity.

Now we have $d-1$ RG invariant parameters
in the gauge theory
and $d-2$ parameters in the string theory.
There is still one more parameter in the gauge theory,
namely the $\beta$-like deformation
given by the baryonic charges (\ref{allbone}).
Substituting (\ref{allbone}) into (\ref{sabz}),
we obtain
\begin{equation}
f^{(2)}
=\sum_{\mu=1}^d \zeta_\mu+c.
\label{f2b}
\end{equation}
If the constraint (\ref{t2cond}) is imposed on
the parameters $\zeta_\mu$,
the first term on the right-hand side of (\ref{f2b}) vanishes,
and this does not give an independent degree of freedom.
To realize the $\beta$-like deformation $f^{(2)}$,
we need to relax the constraint
imposed on the parameters $\zeta_\mu$.
In other words, the marginal deformation $f^{(2)}$
corresponds to a supergravity
field which modifies the constraint (\ref{t2cond}).
We can easily see that the axion field $C$ is such a field.

In \S\ref{strong.sec}, we assumed a vanishing axion field
when we obtained the BPS conditions.
If, instead, we consider a non-vanishing axion field,
the expression for the angle $\varphi$ in the BPS condition (\ref{strongbps}) becomes
\begin{equation}
\varphi\sim\frac{Q(z)}{(N+Q(z)C)g_{\rm str}},
\end{equation}
and
we no longer have the simple relation (\ref{ch-avr}).
This implies that the axion field modifies the constraint (\ref{t2cond}),
and it corresponds to the $\beta$-like deformation (\ref{f2b}).

The correspondence between $f^{(2)}$ and the axion field
can be extended to the correspondence between complex parameters.
With the supersymmetry transformation law of type IIB supergravity,
we can show that the partner of the axion field
is $B_{57}$.
The non-vanishing expectation value of this component of $B_2$
contributes to the phase of the coupling $h_k$
as $h_k\sim e^{-(C+iB_{57})}$ through the coupling of $B_2$ and the string world sheet,
and we obtain the correspondence
\begin{equation}
f^{(2)}\sim C+iB_{57}.
\end{equation}

Having obtained the relations between
background supergravity fields and
the two marginal deformations $f^{(1)}$ and $f^{(2)}$,
the $d-3$ other marginal deformations $f^{(I)}$ ($I=3,\ldots,d-1$)
are naturally matched with the parameters $\zeta_\mu+iM_\mu$.
If we change the parameter $\zeta_\mu$,
the cycle $\mu$ moves on the torus,
and the areas of the faces touching the cycle
change.
This changes the corresponding coupling constants.
If we change the Wilson line $W_\mu$,
the $\theta$-angles associated with the $(N,0)$ faces
touching the cycle are changed
by the interaction term in the brane action
\begin{equation}
\int_{\rm junctions}A^{\rm (NS5)}\wedge F^{\rm (D5)}\wedge F^{\rm (D5)}.
\end{equation}

Summarizing, the relations between exactly marginal deformations in the gauge theory and the degrees of freedom in the brane tiling are given by the following:

\begin{subequations}
\begin{eqnarray}
\mbox{diagonal gauge coupling} & \leftrightarrow & C_{57}+ie^{-\phi}, \label{dilaton}\\
\mbox{$\beta$-like deformation} & \leftrightarrow & C+iB_{57}, \label{axion}\\
\mbox{other $d-3$ deformations} & \leftrightarrow & \zeta_\mu+iW_\mu. \label{zeta}
\end{eqnarray}
\end{subequations}
We should note that there may be a mixing
among these parameters that cannot be captured with the rough analysis given above.

%%%%%%%%%%%%%%%%%%%%%%%%%%%%%%%%%%%%%%%%%%%%%%%%%%%%%%%%%
\section{Conclusions and discussion}\label{conc.sec}
In this paper we have clarified the relation between
the exactly marginal deformations of $\mathcal{N}=1$
superconformal quiver gauge theories and 
the supersymmetry-preserving deformations of the corresponding brane tilings.

First, by analyzing $\beta$-functions for gauge couplings and superpotential couplings,
we showed that there are generically $d-1$ exactly marginal deformations,
where $d$ is the
perimeter of the toric diagram (or the number of legs of the web diagram), 
and have explicitly written down the vanishing linear combinations of $\beta$-functions.
We described a simple method to obtain
the coefficients of the linear combinations
by using the brane tilings.

Next, we treated the brane tiling as a {\it physical} intersecting brane system
and showed that in both the cases of 
large $g_{\rm str}$ and small $g_{\rm str}$
there are $d-3$ degrees of freedom of the deformations of the brane tiling 
which preserve supersymmetry.
In the weak coupling limit, these degrees of freedom correspond to changes
of the coefficients of the Newton polynomial that specifies the holomorphic cycle NS5-brane wraps, whereas in the strong string coupling limit,
these degrees of freedom correspond to the positions of the cycles, 
that is NS5-branes.

We then  combined our results and determined the correspondence between the marginal deformations in the gauge theory and the degrees of freedom of the brane tilings.
Among the $d-1$ degrees of freedom in the gauge theory,
two parameters are identified with background supergravity fields.
One is the ``(complexified) diagonal gauge coupling,''
identified with the dilaton field and the RR two-form field through (\ref{dilaton}),
and the other is the ``$\beta$-like deformation,''
identified with the axion field and the NS-NS two-form field through (\ref{axion}).
The remaining $d-3$ deformations of the gauge theory are
identified with $d-3$ degrees of freedom
of the brane configuration, using the BPS conditions.

%%%%%%%%%%%%%%%%%%%% discussions %%%%%%%%%

There are many open questions which require further investigations.
These are discussed in the following.

First, we found that in the large string coupling limit
there exist BPS brane configurations which have
faces on which the NS5 charge is greater than $1$
or less than $-1$.
Such brane configurations cannot be transformed to
bipartite graphs.
When we consider the duality between quiver gauge theories
and brane tilings, we usually take the weak coupling limit,
and we therfore do not encounter such extra kinds of faces.
However, it may be interesting
to investigate the gauge theories realized on the brane configuration
in the strong coupling limit as an independent problem.

Second, in this paper, we studied only gauge theories with
gauge groups of the same rank.
In the infrared limit, it is believed that such gauge theories flow
into superconformal fixed points.
We can also realize non-conformal gauge theories
with gauge groups of different ranks
by replacing the NS5-branes attached to D5-branes
in the configurations studied in this paper
with $(k,1)$-branes.\cite{Imamura:2006ub} \
It would be interesting to consider
how we can reproduce the
coupling running and the duality cascade
by using brane configurations.
In elliptic models, which are T-dual to brane tilings,
the coupling running is realized as bends
of NS5-branes due to the tension of D4-branes stretched among
them.
A similar mechanism may exist in the case of brane tilings.

Furthermore, in \S \ref{comparison.sec}, we have given only a rough estimation of parameters, and it is desirable to obtain a more complete understanding of the relation between the parameters of gauge theories and those of brane configurations.

Finally, brane-tiling-like descriptions for three-dimensional
$\mathcal{N}=2$ superconformal gauge theories
on M2-branes probing singular Calabi-Yau 4-folds have recently been proposed.
\cite{Lee:2006hw} \
Although we have a much less complete understanding
of theories on M2-branes,
it would be interesting to extend our analysis to that case.

\section*{Acknowledgements}
We thank S.~Lee, Y.~Nakayama, H.~Ooguri and Y.~Tachikawa for valuable comments and discussions. 
We would also like to thank the members of the Yukawa Institute for
Theoretical Physics of Kyoto University.
Discussions during the YITP workshop YITP-W-06-11,
``String Theory and Quantum Field Theory,''
were useful in completing this work.
M.~Y. would like to thank the members of Korea Institute for Advanced Study, where part of this work was completed, for their hospitality.
The research of Y.~I. is supported in part by
a Grant-in-Aid for Scientific Research
(\#17540238)
from the Japan Ministry of Education, Culture, Sports,
Science and Technology.
The research of H.~I. is supported by the Japan Society for the Promotion of Science
(JSPS) under the DC1 program.

\appendix
%%%%%%%%%%%%%%%%%%%%%%%%%
\section{BPS Condition and the Central Charge}\label{bpscond.sec}
In this appendix, we derive the BPS condition for the $(p,q)$-brane
as the $SL(2,{\mathbb Z})$ transformation of that for D5-branes.

The scalar field $V_I$ ($I=1,2$) in type IIB supergravity takes values in
the coset space $SL(2,{\mathbb R})/U(1)$.
This is transformed under the global $SL(2,{\mathbb R})$ and the
local $U(1)$ symmetry operations as
\begin{equation}
V_I\rightarrow V_I'=e^{i\alpha}V_JM^J{}_I,
\label{sl2u1}
\end{equation}
where $M^J{}_I$ is an $SL(2,{\mathbb R})$ matrix.
The following normalization condition is imposed on $V_I$:
\begin{equation}
\Im(V_1^*V_2)=1.
\end{equation}
Due to this condition and the $U(1)$ gauge symmetry,
$V_I$ has two real physical components,
which are often referred to as the dilaton $\phi$ and the axion $C$.
The $U(1)$ invariant physical degree of freedom
is
\begin{equation}\label{tau}
\tau\equiv C+\frac{i}{e^\phi}=\frac{V_2}{V_1}.
\end{equation}
We fix the $U(1)$ gauge by the condition
\begin{equation}
V_1\in{\mathbb R}_+.
\label{u1gauge}
\end{equation}

The SUSY condition for a D5-brane is
\begin{equation}
\Gamma^{012345}\epsilon_i=(\sigma_x)_{ij}\epsilon_j,
\label{d5susy}
\end{equation}
where we have assumed that the D5-brane spreads along
$012345$.
For a D5-brane with a different direction, we need to replace
$\Gamma^{012345}$ by the appropriate gamma matrix.
The parameters $\epsilon_i$ ($i=1,2$) are those
of the supersymmetry transformations.
They are Majorana-Weyl spinors satisfying
\begin{equation}
\Gamma^{11}\epsilon_i=\epsilon_i.
\end{equation}
The $U(1)$ charges of $\epsilon_i$ are $\pm 1/2$, and
the spinors $\epsilon_i$ are transformed under the transformation
(\ref{sl2u1}) as
\begin{equation}
(\epsilon_1\ \epsilon_2)
\rightarrow
(\epsilon_1'\ \epsilon_2')
=(\epsilon_1\ \epsilon_2)
\left(\begin{array}{cc}
\cos(\alpha/2) & \sin(\alpha/2) \\
-\sin(\alpha/2) & \cos(\alpha/2)
\end{array}\right).
\end{equation}
In string theory, the $SL(2,{\mathbb R})$ symmetry of supergravity
is broken to its subgroup $SL(2,{\mathbb Z})$ due to the quantization of
the brane charges.
The $SL(2,{\mathbb Z})$ transformation of the $5$-brane charges $Q^I$
is given by
\begin{equation}
Q^I\rightarrow Q^{'I}=M^I{}_JQ^J.
\end{equation}

The SUSY condition for the $(p,q)$-branes can be obtained
as the $SL(2,{\mathbb Z})$ transformation of that of D5-branes (\ref{d5susy}).
A D5-brane is transformed to a $(p,q)$-brane
by the $SL(2,{\mathbb Z})$ transformation
\begin{equation}
M^I{}_J=\left(\begin{array}{cc}
p & r \\
q & s
\end{array}\right),\quad
ps-qr=1.
\end{equation}
We also need to perform the local $U(1)$ transformation $e^{i\varphi}$,
with
\begin{equation}
\varphi=\arg(p+q\tau),
\end{equation}
in order to maintain the gauge (\ref{u1gauge}).
This $U(1)$ transformation rotates the SUSY transformation parameters $\epsilon_i$,
and this transformation changes the SUSY condition (\ref{d5susy}) into
\begin{equation}
\Gamma^{012345}\epsilon_i=\epsilon_j(\sin\varphi\sigma_z+\cos\varphi\sigma_x)_{ji}.
\label{susy10}
\end{equation}
This is the BPS condition for $(p,q)$-branes
spread along $012345$.

In this paper,
we consider $5$-branes spread along $0123$ and
wrapped on a two-dimensional surface in $4567$ space.
To facilitate the study of such configurations,
we decompose the gamma matrices as
\begin{subequations}
\begin{eqnarray}
\Gamma^\mu&=&\gamma^\mu\otimes{\bf1}_4\otimes{\bf1}_2,\\
\Gamma^a&=&\gamma^5\otimes\gamma^a\otimes{\bf1}_2,\\
\Gamma^8&=&\gamma^5\otimes\gamma^5\otimes\sigma_x,\\
\Gamma^9&=&\gamma^5\otimes\gamma^5\otimes\sigma_y,\\
\Gamma^{11}&=&\gamma^5\otimes\gamma^5\otimes\sigma_z,
\end{eqnarray}
\end{subequations}
where $\gamma^\mu$ and $\gamma^a$ are gamma matrices
in the $0123$ and $4567$ spaces, respectively.
Correspondingly, the SUSY parameter
$\epsilon_i$ for each $i$ can be decomposed into
four four-dimensional Weyl spinors of positive chirality
and their complex (Majorana) conjugates.
Let $\epsilon^A{}_i$ be the Weyl spinors of positive chirality,
where $A=1,2,3,4$ is the Dirac spinor index in the $4567$ space.
The quantity $\epsilon^A{}_i$ for each $A$ and $i$ is a chiral spinor in $0123$ space.
With this decomposition of the gamma matrices and spinors,
the SUSY condition (\ref{susy10}) becomes
\begin{equation}
\frac{i}{2}S^{ab}(\gamma_{ab})^A{}_B\epsilon^B{}_i
=\epsilon^A{}_j(\sin\varphi\sigma_z+\cos\varphi\sigma_x)_{ji}.
\label{sustcond4d}
\end{equation}
Here, we have introduced the anti-symmetric tensor $S^{ab}$
specifying the tangent space of the
$5$-brane worldvolume in $4567$ space,
which is defined by
\begin{equation}
S^{ab}=\int_T dx^a\wedge dx^b,
\end{equation}
where $T$ is a tangent plane of unit area.
For the gamma matrices $\gamma^{ab}$ in the $4567$ space,
we use the following convention:
$$
i\gamma^{45}=\left(\begin{array}{cc} \sigma_z \\ & \sigma_z \end{array}\right),\quad
i\gamma^{76}=\left(\begin{array}{cc} \sigma_z \\ & -\sigma_z \end{array}\right),
$$
$$
i\gamma^{46}=\left(\begin{array}{cc} \sigma_x \\ & \sigma_x \end{array}\right),\quad
i\gamma^{57}=\left(\begin{array}{cc} \sigma_x \\ & -\sigma_x \end{array}\right),
$$
\begin{equation}
i\gamma^{47}=\left(\begin{array}{cc} \sigma_y \\ & \sigma_y \end{array}\right),\quad
i\gamma^{65}=\left(\begin{array}{cc} \sigma_y \\ & -\sigma_y \end{array}\right).
\label{explicitgamma}
\end{equation}

In this paper, we are interested in brane configurations
which preserve specific supersymmetries.
These supersymmetries are the unbroken supersymmetries
in the system consisting of NS5-branes along the 45 and 67 directions
and D5-branes along the 57 directions.
For simplicity, we assume a vanishing axion field.
This implies that $\varphi=\pi/2$ for NS5-branes.
(For D5-branes the angle is always $\varphi=0$, by definition.)
The two kinds of NS5-branes give
\begin{equation}
i(\gamma^{45})^A{}_B\epsilon^B{}_i
=i(\gamma^{67})^A{}_B\epsilon^B{}_i
=\epsilon^A{}_j(\sigma_z)_{ji}.
\label{ns5susyc}
\end{equation}
Using the explicit expressions for the gamma matrices,
we can easily show from (\ref{ns5susyc}) that the bottom two rows of the
$4\times2$ matrix $\epsilon^A{}_i$ vanish.
Let $\epsilon_U$ be the upper half of the matrix $\epsilon^A{}_i$.
Then, we can rewrite (\ref{ns5susyc}) as
\begin{equation}
\sigma_z\epsilon_U=\epsilon_U\sigma_Z,
\end{equation}
and this implies that the $2\times 2$ matrix $\epsilon_U$ is diagonal.
Therefore, there are two independent Weyl spinors in $\epsilon_U$.
These are the parameters of the four-dimensional ${\cal N}=2$ supersymmetry
preserved by the NS5-branes.

The SUSY condition for the D5-branes is
\begin{equation}
\frac{i}{2}S_{\rm D5}^{ab}(\sigma_{ab})^A{}_B\epsilon^B{}_i=\epsilon^A{}_j(\sigma_x)_{ji},
\label{d5susyc}
\end{equation}
where $\sigma^{ab}$ is the positive chirality part of $\gamma^{ab}$.
For this condition not to break all the supersymmetries
and to preserve ${\cal N}=1$ supersymmetry,
the matrix $S_{\rm D5}=(1/2)S_{\rm D5}^{ab}\sigma_{ab}$ must be skew-diagonal.
In other words, the two diagonal components of $S_{\rm D5}$ must vanish.
If this condition is satisfied,
the matrices $S$ take the form
\begin{equation}
S=\left(\begin{array}{cc}
0 & Z \\
Z^* & 0
\end{array}\right).
\label{matrixs}
\end{equation}
Here, the component $Z$ satisfies $|Z|=1$, and
its phase determines the unbroken supersymmetry
as
\begin{equation}
\epsilon^2{}_2=Z\epsilon^1{}_1.
\end{equation}
The parameter $Z$, which depends on the direction of the
D5-brane, is merely (the density of) the
central charge in the ${\cal N}=2$ superalgebra,
which is carried by the D5-branes.
This can be shown by noting that
the commutator of local supersymmetry transformations
generates a gauge transformation of the RR $6$-form potential
field, which couples to D5-branes:
\begin{equation}
\delta(\epsilon)\delta(\epsilon')C_6-\delta(\epsilon')\delta(\epsilon)C_6
=d\Lambda_5+\cdots,\quad
\Lambda_5=\frac{1}{4e^\phi}(\sigma_x)_{ij}(\ol\epsilon_i\Gamma_{[5]}\epsilon'_j).
\label{commsusy}
\end{equation}
Here, $\Gamma_{[5]}$ is the matrix-valued five-form defined by
$(1/5!)\Gamma_{M_1\cdots M_5}dx^M\wedge\cdots\wedge dx^{M_5}$.
The blobs in (\ref{commsusy}) represent
the field dependent terms induced by the diffeomorphism which
emerges in the commutator.
We can read off the central charge of the D5-brane
by comparing (\ref{commsusy}) and
the relation
\begin{equation}
\int\Lambda_5=\frac{1}{2}\Re\Big(VZ(\epsilon^1{}_1\epsilon^2{}_2)\Big)
+\cdots,
\end{equation}
where the integral on the left-hand side is taken over
a time slice of the world volume of the D5-brane.
We factor out the divergent volume factor $V$
on the right-hand side, and $Z$ is the central charge of branes
occupying a unit volume in $123$ space.
Using explicit forms of the $\gamma$ matrices,
we can show that the central charge is
the component $Z$ in the matrix (\ref{matrixs}).

The SUSY condition for the D5-branes along the 57 directions is
\begin{equation}
i(\gamma^{57})^A{}_B\epsilon^B{}_i=\epsilon^A{}_j(\sigma_x)_{ji}.
\label{d5susyc-2}
\end{equation}
The conditions (\ref{ns5susyc}) and (\ref{d5susyc-2}) imply that this $2\times2$ matrix $\epsilon_U$ commutes
with $\sigma_z$ and $\sigma_x$, respectively,
and therefore $\epsilon_U$ is proportional to the unit matrix.
Explicitly,
the matrix $\epsilon$ is given by
\begin{equation}
\epsilon^A{}_i=\left(\begin{array}{cc}
\epsilon & 0 \\
0 & \epsilon \\
0 & 0 \\
0 & 0
\end{array}\right).
\label{unbsusy}
\end{equation}
The four-dimensional Weyl spinor $\epsilon$ in
(\ref{unbsusy}) is the parameter of unbroken ${\cal N}=1$ SUSY.

We are interested in brane configurations
which do not break the ${\cal N}=1$ SUSY.
By substituting 
(\ref{unbsusy}) into
(\ref{sustcond4d}),
we obtain
\begin{equation}
t_1^a t_2^b (i\sigma^{ab})
=\sin\varphi\sigma_z+\cos\varphi\sigma_x.
\label{pqcons}
\end{equation}
This equation reduces to (\ref{strongbps}) when $\varphi$ is small.

\end{document}